\documentclass[reqno,11pt]{article} 

\usepackage{amsmath,amsthm,amssymb,amscd,amstext,amsfonts}
\usepackage{mathtools}
\usepackage{mathrsfs}
\usepackage{thmtools}
\usepackage{latexsym}
\usepackage{verbatim}
\usepackage{framed}
\usepackage{graphicx}
\usepackage{stmaryrd}
\usepackage{enumerate}
\usepackage{fullpage}
\usepackage{bm}
\usepackage{color}
\usepackage{hyperref}
\usepackage{url}
\usepackage{physics}
\usepackage{multicol}
\usepackage{multirow}
\usepackage{tikz-cd}
\usetikzlibrary{decorations.pathmorphing}   
\usetikzlibrary{arrows}
\usepackage{makeidx}
\usepackage{makecell}		

\makeindex

\usepackage{caption} 
\usepackage[basic]{complexity}    

\newtheorem{theorem}{Theorem}

\newtheorem{corollary}[theorem]{Corollary}
\newtheorem{lemma}[theorem]{Lemma}
\newtheorem{proposition}[theorem]{Proposition}
\newtheorem{definition}[theorem]{Definition}
\newtheorem{example}[theorem]{Example}

\newtheorem{remark}[theorem]{Remark}

\DeclareMathOperator{\Nb}{Nb}   
\DeclareMathOperator{\GAL}{GAL}    
\DeclareMathOperator{\BW}{BW}    
\DeclareMathOperator{\PG}{PG}    

\DeclarePairedDelimiter\ceil{\lceil}{\rceil}   
\DeclarePairedDelimiter\floor{\lfloor}{\rfloor}

\newcommand{\inp}[2]{\left\langle#1,#2\right\rangle}            
 
\newcommand{\mult}{\mathsf{mult}}
\newcommand{\nrows}{\mathsf{nrows}}

\newcommand{\ccmat}{\mathsf{ccm}}
\newcommand{\nccmat}{\mathsf{nccm}}

\DeclareMathOperator{\AND}{AND}
\DeclareMathOperator{\OR}{OR}

\DeclareMathOperator{\EQ}{EQ}                       
\DeclareMathOperator{\SEQ}{SEQ}                       
\DeclareMathOperator{\PARITY}{PARITY}

\DeclareMathOperator{\TEP}{{TEP}}          
\DeclareMathOperator{\GEN}{{GEN}}          
\DeclareMathOperator{\BRS}{{BRS}}          
\DeclareMathOperator{\TS}{{TS}}          
\DeclareMathOperator{\ISA}{ISA}   
\DeclareMathOperator{\NAND}{NAND}  

\DeclareMathOperator{\CC}{CC}        
\DeclareMathOperator{\NCC}{NCC}        
\DeclareMathOperator{\BP}{BP}        
\DeclareMathOperator{\nBP}{NBP}        
\DeclareMathOperator{\OBDD}{OBDD}        
\DeclareMathOperator{\BPone}{\BP_1}        
\DeclareMathOperator{\nBPone}{\nBP_1}        
\DeclareMathOperator{\DNFS}{DNFSize}        
\DeclareMathOperator{\CNFS}{CNFSize}        
\DeclareMathOperator{\DTS}{DTSize}        
\DeclareMathOperator{\clique}{CLIQUE}        

\DeclareMathOperator{\longSym}{Sym}    

\newcommand{\F}{\mathbb{F}}

\newcommand{\N}{\mathbb{N}}

\renewcommand{\R}{\mathbb{R}}  

\newcommand{\cC}{\mathcal C}

\newcommand{\cF}{\mathcal F}

\renewcommand{\cL}{\mathcal L}  

\renewcommand{\cP}{\mathcal P}  

\allowdisplaybreaks

\begin{document}
\title{Perspective on complexity measures targetting read-once branching programs}
\author{Yaqiao Li\footnote{Concordia University, yaqiao.li@concordia.ca}, 
Pierre McKenzie\footnote{Universit{\'e} de Montr{\'e}al, mckenzie@iro.umontreal.ca}}

\maketitle

\begin{abstract}
A model of computation for which reasonable yet still incomplete lower bounds are known is the read-once branching program. Here variants of complexity measures successful in the study of read-once branching programs are defined and studied. Some new or simpler proofs of known bounds are uncovered. Branching program resources and the new measures are compared extensively.
The new variants are developed in part in the hope of tackling read-$k$ branching programs for the tree evaluation problem~\cite{cook2012pebbles}. Other computation problems are studied as well. In particular, a common view of a function studied by G\'{a}l~\cite{gal1997simple} and a function studied by Bollig and Wegener~\cite{bollig1998very} leads to the general combinatorics of blocking sets. Technical combinatorial results of independent interest are obtained. New leads towards further progress are discussed.
An exponential lower bound for non-deterministic read-$k$ branching programs for the GEN function~\cite{JonesGEN} is also derived, independently from the new measures.
\end{abstract}

\tableofcontents

\section{Introduction}

Proving lower bounds on the resources needed to perform a computation often relies on confronting the computation model with complexity measures capturing its combinatorics.
For example, partitions into rectangles and ranks of matrices offer viewpoints on the two-party communication complexity of a boolean function. As a rule of thumb, the more viewpoints are available, the more successful our study of the model is.

Polynomial size branching programs (bps) have long been known to capture logarithmic space. Much effort was devoted to their study and lower bounds on restricted bps abound (see~\cite{wegener2000branching}). But even in the context of as severe a restriction as \emph{read-once}, or even \emph{ordered read-once}, no lower bound method applies naturally to all hard functions in $\L$, in $\NL$, or even in $\P$ for that matter, despite the belief that $\P$-hard functions require exponential size \emph{unrestricted} bps. Read-once bps and their variants remain themselves to this day an object of study in connection with derandomization (e.g.~\cite{forbes2018pseudorandom,tashma2021PrgsAgainstReadOnce}) and proof complexity (e.g.~\cite{Tseitin2017,sofronovaSokolov2021boundedRepetition}).  

In this work we first define two types of complexity measures inspired by known lower bounds for read-once bps. The first type (Section~\ref{sec:def-subf}) derives from counting subfunctions of the function being computed. The second type (Sections~\ref{sec:def-cover}, \ref{sec:def-partition}) exploits the variable partition model in communication complexity and leads to a framework tersely described as max-min communication complexity (Section~\ref{sec:CC}).

The new measures are then compared with each other and with read-once deterministic and nondeterministic BP sizes (Figure~\ref{fig:relation} and Table~\ref{table:compare}). Several separations there follow from known (or adaptations of known) upper and lower bounds for the functions defined and grouped in Section~\ref{sec:funcs}. But a perspective on the combinatorics of read-limited bps emerges and in subsequent sections we pick up on some of the threads that arise.

In Section~\ref{sec:TEP} we consider the tree evaluation problem, proposed in~\cite{cook2012pebbles} as a candidate to separate $\L$ from larger classes. A read-once BP size lower bound for $\TEP$  is known~\cite{iwama2018read}.  In fact, our initial motivation for considering subfunctions counting lower bound measures for read-once bps was to use them to give an alternative proof for a read-once BP size lower bound for $\TEP$, in hoping that the alternative proof can be generalized to the read-$k$ case. However, we show that a weak form of subfunctions counting (measure $\widehat{S}$) will not suffice for this purpose. Our stronger measure $S$ implies lower bounds on ordered read-once BP size in general (Corollary \ref{cor:S-OBDD}). We provide an incomplete report on $S(\TEP)$.

In Section~\ref{sec:Tseitin}, we use our weak covering measure (measure $\widehat{C}$) to give an alternative and simpler proof of a lower bound on the size of Tseitin formulas, used in~\cite{Tseitin2019} as a tool to obtain nondeterministic read-once BP lower bounds on the satisfiability problem for such formulas.

In Section~\ref{sec:Gal} we cast two functions known to require large read-once bps, namely GAL defined in~\cite{gal1997simple} from projective geometry and BW defined in~\cite{bollig1998very} from representing numbers in a prime basis, as problem instances in a common regular $K_{2,2}$-free bipartite graph.  GAL and BW are known to have small CNF size and DNF size respectively, yet no function with small weight (i.e., small DNF and small CNF sizes \emph{combined}) is known to require large read-once bps. We observe that weight and \emph{ordered} read-once BP size are provably unrelated.  Then we extend the GAL lower bounds to the bipartite graph setting by means of our measures $\widehat{S}$ and $\widehat{C}$. Our analysis suggests the need to better understand the properties of blocking sets in $\F_p^2$, raising several questions of a purely combinatorial nature. We make some observations in that direction (such as Lemma~\ref{lem:intersecting-points-property} describing an elegant property of intersecting points in $\F_p^2$).

In Section~\ref{sec:GEN} we record an exponential nondeterministic read-$k$ BP lower bound for the $\P$-complete GEN function.  This is obtained independently from our measures, by merely exhibiting a read-once reduction from the BRS function defined in~\cite{BRS1993}. 

In Section~\ref{sec:openproblem} we take stock and highlight several open questions arising from this work, some of which are of independent interest from the viewpoint of combinatorics alone.

\section{Measures, branching programs and common functions}
\label{sec:def}

This section introduces notation, defines our measures, defines max-min complexity, recalls the definitions of branching programs and finally collects the definitions of several known functions whose complexities are at stake in the paper.

\subsection{Notation} \label{sec:notation}

The set $\{1,\ldots,k\}\subseteq \N$ 
is denoted $[k]$. The operations $\bigvee$ and $\bigwedge$ denote the Boolean OR and AND on $\{0,1\}$. The operations $+$ and $\cdot$ are the usual sum and product on $\N$ or $\R$. When $f,g:D\rightarrow E$ and $\mathsf{op}$ is a binary operation on $E$, $f\ \mathsf{op}\ g$ is the function defined by $(f\ \mathsf{op}\ g)(x)= f(x)\ \mathsf{op} \ g(x)$.

Let $f:\{0,1\}^n\rightarrow \N$.
Let
$A=\{i_1,\ldots,i_\ell\}\subseteq [n]$,  
$\overline{A}=\{j_1,\ldots,j_{n-\ell}\}\subseteq [n]$,  
$i_1<\cdots < i_\ell$ and $j_1<\cdots <j_{n-\ell}$.
We define $\pi_A\in\longSym(n)$ as the permutation $$
\left(
\begin{array}{cccccccc}
1 & 2 & \cdots & \ell & \ell+1 & \ell+2 & \cdots & n\\
i_1 & i_2 & \cdots & i_\ell & j_1 & j_2 & \cdots & j_{n-\ell}\\
\end{array}
\right)$$ 
and write $f_A$ for the $2^\ell \times 2^{(n-\ell)}$ matrix defined by
\begin{align*}
	f_A[\ (x_{\pi_A(1)}, \ldots, x_{\pi_A(\ell)}),\ (x_{\pi_A(\ell+1)}, \ldots, x_{\pi_A(n)})\ ] = f(x_1,\ldots,x_n).
\end{align*}
 Any $\alpha \in \{0,1\}^A$ then identifies a row of $f_A$ and prescribes the subfunction $f_{A,\alpha}: \{0,1\}^{\overline{A}} \to \N$ of $f$ given by 
    $f_{A,\alpha}(y) = f_A(\alpha,y)$.

\subsection{Measures from counting subfunctions: $S$ and $\widehat{S}$}   \label{sec:def-subf}

\begin{definition}   \label{def:subfuncs}
Let $f: \{0,1\}^n \to \N$ and $A \subseteq [n]$.  Let $\nrows(f_A)$ denote the number of distinct rows of the matrix $f_A$ and $\mult(f_A)$ the number of occurrences of a most frequent row. Define
    \begin{align}   \label{eq:s}
        S(f) &= \max_{1 \le k\le n} \min_{A \subseteq [n], |A| = k} \nrows(f_A),\\
           \label{eq:hat-s}
        \widehat{S}(f) &= \max_{1 \le k \le n} \min_{A \subseteq [n], |A| = k} \frac{2^k}{\mult(f_A)}.
    \end{align}
\end{definition}

Consider the multiset of subfunctions $\{f_{A,\alpha}: \alpha \in \{0,1\}^A \}$. This is the same as the multiset of rows of the matrix $f_A$. Then,  $\nrows(f_A)$ is the number of distinct subfunctions, and $\mult(f_A)$ is the \emph{multiplicity} of this multiset. For example, the notion of an \emph{$m$-mixed} Boolean function (see \cite{mixed}) is captured via $\mult(f_A)$ as follows: $f$ is \emph{$m$-mixed} if $\mult(f_A) =1$ for every $A\subseteq [n]$ of size $|A|= m$. 

\begin{lemma}   \label{lem:mixed-to-S-hat}
    If $f$ is $m$-mixed, then
        $\widehat{S}(f) \ge 2^m$.
\end{lemma}

\begin{remark}  \label{rmk:S-on-D-ary}
The measures $S(f)$ and $\widehat{S}(f)$ can be generalized to non-Boolean functions 
    $f: D^n \to R$, by leaving
\eqref{eq:s} unchanged and replacing in
\eqref{eq:hat-s} the numerator $2^k$  with $|D|^k$. 
\end{remark}

A lower bound on OBDD size (Section~\ref{sec:OBDD}) will involve a min-max ``dual'' to the $S$ measure. For
$\sigma\in\longSym(n)$ and $1\le k \le n$, let 
    $\sigma([k])$ denote the set
$\{\sigma(1), \ldots, \sigma(k)\}$ of size $k$.
With this notation,  
    $S(f) = \max_{1 \le k\le n} \min_{|A| = k, A \subseteq [n]} \nrows(f_A)
        = \max_{1 \le k\le n} \min_{\sigma \in \longSym(n)} \nrows(f_{\sigma([k])})$.

\begin{definition}   \label{def:dual-S}
Define the dual form of $S$, denoted by $S^*$, as
    \begin{equation}   \label{eq:dual-S}
        S^*(f) = \min_{\sigma \in \longSym(n)} \max_{1 \le k\le n}  \nrows(f_{\sigma([k])}).
    \end{equation}
\end{definition}

By definition, $S^*(f) \ge S(f)$.

\subsection{Rectangles}   \label{sec:def-rectangle}

The measures to be defined in Sections~\ref{sec:def-cover} and~\ref{sec:def-partition} build on the notion of a rectangle, standard for functions with a Boolean range but in need of a definition for functions with range $\N$.

\begin{definition}   \label{def:matrix-rectangle}
A matrix $M\in \N^{k\times \ell}$ is a \emph{rectangle} if $M(i,j)= g_i \cdot h_j$ for some $g\in \{0,1\}^k$ and $h\in \N^\ell$. 
\end{definition}

Note the provenance of $g$ in Definition~\ref{def:matrix-rectangle} (drawing $g$ from $\N^k$ would make sense as well but we only have need for Boolean $g$s in our applications). Note further that when $M$ is Boolean, a rectangle as defined is a rectangle in the standard sense~\cite[Def 1.12]{kushilevitz1997communication}.

\begin{definition} \label{def:function-rectangle}
Let $f:\{0,1\}^n \to \N$, $A\subseteq [n]$ and $k\in [n]$. Then	
	$f$ is declared an $A$-\emph{rectangle} if the matrix $f_A$ is a rectangle, and 
	declared a $k$-rectangle if it is an $A$-rectangle for some $A$ of size $k$.
\end{definition}

\begin{example}
Let $n$ be an integer multiple of $4$. Consider the equality function $g: \{0,1\}^{n} \to \{0,1\}$ defined by $g(x_1, \ldots, x_n) = 1$ if and only if $x_i = x_{n/2 + i}$ for every $1 \le i \le n/2$. With $A = \{1,\ldots, n/2\}$, $g_A$ is simply the identity matrix, hence is \emph{not} an $A$-rectangle. Consider $A' = \{1,\ldots, n/4, n/2+1, \ldots, 3n/4\}$. It is easy to see that for every $a\in \{0,1\}^{A'}, b\in \{0,1\}^{\overline{A'}}$, $g(a,b) = g_1(a) \land g_2(b) = g_1(a) \cdot g_2(b)$, where $g_1$ and $g_2$ are both the equality function on $n/2$ bits. Hence $g$ is an $A'$-rectangle. As $|A'| = n/2$, $g$ is an $n/2$-rectangle. 
\end{example}

\subsection{Measures from covering: $C$ and $\widehat{C}$}   \label{sec:def-cover}

The measures defined in this section relate to nondeterministic communication complexity, hence apply solely to functions with a Boolean range. 

\begin{definition} \label{def:covering}
Let $f:\{0,1\}^n\rightarrow \{0,1\}$, $A\subseteq [n]$ and $k\in [n]$. Define
	\begin{align}\label{eqn:covering}
		C(f,A) &= \min \left\{r: f \text{ is the $\bigvee$ of $r$ functions that are $A$-rectangles}\right\}\\
		C(f,k) &= \min \left\{r: f \text{ is the $\bigvee$ of $r$ functions that are $k$-rectangles}\right\}\\
		C(f) &= \max_{k\in [n]}\min_{A\subseteq [n], |A|=k} C(f,A)\\
		\widehat{C}(f) &= \max_{k\in [n]}\ C(f,k).
	\end{align}
\end{definition}

We will refer to $C(f)$ as the cover number of $f$ and to $\widehat{C}(f)\leq C(f)$ as its weak cover number. Justification for naming $C(f)$ in this way comes from observing that when $A\subseteq [n]$, $C(f,A)$ is the measure $C^1(f_A)$ defined in~\cite[Def 2.1]{kushilevitz1997communication} as the minimal number of $1$-monochromatic rectangles needed to cover the ones in the matrix $f_A$. Of course $\widehat{C}(f)$ is a new measure to be studied. We note that the appellation ``cover number of $f$'' in \cite{kushilevitz1997communication} applies in the fixed partition communication model and differs from our $C(f)$ here.

\subsection{Measures from partitioning: $P$ and $\widehat{P}$}   \label{sec:def-partition}

Here we allow functions with a non-Boolean range. We say that $f:D\to \N$ and $g:D\to \N$ are orthogonal if $f \cdot g$ is the zero function.

\begin{definition} Let $f:\{0,1\}^n\rightarrow \N$, $A\subseteq [n]$ and $k\in [n]$. \label{def:partition-numbers}
		\begin{align}\label{eq:def-partition-num}
		P^+(f,A) &= \min \left\{r: f \text{ is the integer sum of $r$ pairwise orthogonal $A$-rectangles}\right\}\\
		P^+(f,k) &= \min \left\{r: f \text{ is the integer sum of $r$ pairwise orthogonal $k$-rectangles}\right\}\\
		P^+(f) &= \max_{k\in [n]}\min_{A\subseteq [n], |A|=k} \ P^+(f,A)\\
		\widehat{P^+}(f) &= \max_{k\in [n]}\ P^+(f,k)
		\intertext{and further when $\text{Im}(f) \subseteq \{0,1\}\subseteq\N$,}
		P(f) &= \max_{k\in [n]}\min_{A\subseteq [n], |A|=k} (P^+(f,A) + P^+(\neg f,A))\\
		\widehat{P}(f) &= \max\left\{ \widehat{P^+}(f), \widehat{P^+}(\neg f)\right\}.
	\end{align}
\end{definition}

Again here we will refer to $P^+(f)$ and to $\widehat{P^+}(f)\leq P^+(f)$ (and  to $P(f)$ and to $\widehat{P}(f)\leq P(f)$ when applicable) respectively as  the partition number and the weak partition number of $f$. Justification for this naming comes from the partition number $C^D(M)$~\cite[Def 2.1]{kushilevitz1997communication} of a Boolean matrix $M$, defined as the minimum number of monochromatic rectangles needed to partition $M$. When $\text{Im}(f) \subseteq \{0,1\}$, $P^+(f,A)$ refers to partitioning the ones of the Boolean matrix $f_A$, so $P^+(f,A)+P^+(\neg f,A)= C^D(f_A)$.

\begin{remark}
The partition number $P(f)$ is investigated under the name ``rectangle complexity'' in 
\cite{sauerhoff2003approximation}, where  errors are allowed in representing $f$ using rectangles.
\end{remark}

\subsection{max-min Communication complexity}   \label{sec:CC}

Let $M$ be a Boolean matrix. Let $\ccmat(M)$ denote its deterministic communication complexity (i.e., measure $D$ in~\cite[Def 1.2]{kushilevitz1997communication}) and $\nccmat(M)$ denote the nondeterministic communication complexity of $M$ (i.e. measure $N^1$ in~\cite[Def 2.3]{kushilevitz1997communication}).

\begin{definition}   \label{def:CC-NCC}
Let $f: \{0,1\}^n \to \{0,1\}$. 
Define the \emph{max-min communication complexity of $f$}, denoted by $\CC(f)$, to be
\begin{equation}   \label{eq:def-max-min-CC}
\CC(f) =  \max_{1\le k \le n} \min_{A \subseteq [n], |A| = k} \ccmat(f_A)
\end{equation}
and the \emph{max-min nondeterministic communication complexity of $f$}, denoted by $\NCC(f)$, to be
\begin{equation}   \label{eq:def-max-min-NCC}
\NCC(f) =  \max_{1\le k \le n} \min_{A \subseteq [n], |A| = k} \nccmat(f_A).
\end{equation}
\end{definition}

\begin{remark}
The quantity $\min_{A \subseteq [n], |A| = n/2} \ccmat(f_A)$ corresponds to the notion  $D^{\text{best}}(f)$ in \cite{kushilevitz1997communication}. By definition, 
    \begin{equation} \label{eq:NCC-C}
        \NCC(f) = \log C(f). 
    \end{equation}
\end{remark}

Note that $\CC(f)$ seamlessly generalizes to non-Boolean functions $f: D^n \to \{0,1\}$.
Reassuringly $\CC(f)$ and $\NCC(f)$ in the Boolean case behave as we expect:
\begin{lemma}
Let $f: \{0,1\}^n \to \{0,1\}$. Then
  \begin{equation}   \label{eq:CC-ub}
        \NCC(f) \le \CC(f) \le n/2 + 1,
    \end{equation}
    \begin{equation}   \label{eq:P-and-CC}
        \log P(f) \le \CC(f) \le O(\log^2 P(f)).
    \end{equation}
\end{lemma} 
\begin{proof}
The left of (\ref{eq:CC-ub}) holds since $\nccmat\leq \ccmat$. The right of (\ref{eq:CC-ub}) holds because $\ccmat(f_A)\leq 1+\min\{|A|,n-|A|\}$. As to (\ref{eq:P-and-CC}), recall that $P^+(f,A) + P^+(\neg f,A)= C^D(f_A)$, which is at most $2^{\ccmat (f_A)}$ by~\cite[Prop 2.2]{kushilevitz1997communication}, so applying $\max_k \min_A$ yields $P(f)\le 2^{\CC(f)}$. Finally, any Boolean matrix of the form $f_A$ satisfies $\ccmat(f_A)\le c\cdot\log^2P^+(f,A)$ for some constant $c$ (see~\cite[Ex 1.1]{rao2020communication}). Hence $\ccmat(f_A)\leq c\cdot\log^2(P^+(f,A)+P^+(\neg f, A))$, so that applying $\max_k \min_A$ yields $\CC(f) \le c\cdot \log^2 P(f)$.
\end{proof}

\subsection{Branching programs}

A deterministic (binary) branching program (BP for short) is a directed acyclic graph (DAG for short) with a unique source node and two sink nodes (one sink node labelled by $1$, while the other sink node labelled by $0$). Each non-sink node has outdegree $2$, the node is labelled by a variable $x_i$ for some $i \in [n]$, one of the two out-edges of the node is labelled by $x_i=0$ and the other is labelled by $x_i=1$. Every $x \in \{0,1\}^n$ defines a unique source-to-sink path in a BP. A BP computes a Boolean function $f: \{0,1\}^n \to \{0,1\}$ if the unique source-to-sink path for $x$ ends at the sink with label $f(x)$. The size of a BP is defined to be the number of nodes. Let $\BP(f)$ denote the smallest deterministic BP size computing $f$ correctly.
A BP is called syntactic \emph{read-$k$} if on every source-to-sink path, each variable $x_i$ appears at most $k$ times. Let $\BP_k(f)$ denote the minimal size among all read-$k$ BPs that  compute $f$.

A nondeterministic BP (NBP for short) is a DAG with a unique source node and a unique sink node (the sink node is labelled by $1$). Each non-sink node has outdegree at most $2$. The non-sink nodes have no labels. Each edge is either labelled by $x_i=0$, or by $x_i=1$, or has no label. An edge with no label will pass through all inputs. An NBP computes a Boolean function $f: \{0,1\}^n \to \{0,1\}$ if for every input $x$ such that $f(x)=1$, there exists at least one source-to-sink consistent path with $x$. The size of an  NBP is defined to be the number of labelled edges.  An NBP is called syntactic \emph{read-$k$} if on every source-to-sink path, each variable $x_i$ appears at most $k$ times. Let $\nBP_k(f)$ denote the minimal size among all read-$k$ NBPs that compute $f$.

An OBDD (ordered binary decision diagram, a.k.a., oblivious read-once branching program) is a read-once BP with the following property: there is a permutation $\pi$ on $[n]$, so that variables on every source-to-sink path follow the order $\pi$. Specifically, if $x_j$ appears after $x_i$ in a source-to-sink path, then $\pi^{-1}(j) > \pi^{-1}(i)$. Let $\OBDD(f)$ denote the minimal size among all OBDDs that compute $f$. 

By definition, for every Boolean function $f$, $\OBDD(f) \ge \BP_1(f) \ge \nBP_1(f)$. 

\subsection{Roster of functions}    \label{sec:funcs}

Here we define all the functions that will be used.

\begin{itemize}
    \item \emph{The equality function} $\EQ_n:  \{0,1\}^n \times \{0,1\}^n \to \{0,1\}$ is defined by $\EQ_n(x,y) = 1$ if and only if $x=y$. 

	\item \emph{The shifted equality function} $\SEQ_n: \{0,1\}^n \times \{0,1\}^n \times [n] \to \{0,1\}$, is defined by $\SEQ_n(x,y,i) = 1$ if and only if $x_j = y_{t_j}$ for all $j=1,\ldots, n$, where $t_j \in [n]$ and $t_j  \equiv j+i-1 \mod n$. That is, $\SEQ_n(x,y,i)$ computes the equality function on $x$ and $y$, where bits of $y$ are shifted by $i-1$ to the left.

	\item \emph{The parity function}  $\PARITY_n: \{0,1\}^n \to \{0,1\}$ is defined by $\PARITY_n(x) = 1$ if and only if the number of $1$s in $x$ is odd.

	\item \emph{The exact half clique function}  $\clique_{n,n/2}: \{0,1\}^{n(n-1)/2} \to \{0,1\}$. Given an input $x \in \{0,1\}^{n(n-1)/2}$ as a graph on $n$ vertices, $\clique_{n,n/2}(x) = 1$ if and only if the graph $x$ contains exactly a clique of size $\ceil{n/2}$ and $\floor{n/2}$ isolated vertices.

	\item \emph{The pointer function} $\pi_n: \{0,1\}^n \to \{0,1\}$. To define the pointer function we firstly define the $(\OR \circ \AND)_{m^2}: \{0,1\}^{m^2} \to \{0,1\}$ as $(\OR \circ \AND)_{m^2}(y) = \lor_{i=1}^{m} \land_{j=1}^m y_{(i-1)m+j}$. Now write $n = \log n \times \frac{n}{\log n}$ and partition the input $n$-bits into $\log n$ blocks $x = (x_1,..., x_{\log n})$ where each $x_i \in \{0,1\}^{n/ \log n}$ for $i=1,\ldots, \log n$. Let $z_i = (\OR \circ \AND)_{n/\log n}(x_i)$. Let $0 \le z \le n-1$ denote the unique integer represented by the $\log n$ bits $z_1 \cdots z_{\log n}$. Then, define $\pi_n(x) = x_{z+1}$, i.e., the $(z+1)$-th bit in $x$. 

	\item \emph{The Tree Evaluation Problem (TEP)}. 
		Let 
			$(\TEP,1): [k] \to [k]$ 
		be defined as 
			$(\TEP,1)(x) = x$. 
		For $h\ge 2$, let 
			$(\TEP,h): [k]^{n_h} \to [k]$ 
		denote the Tree Evaluation Problem (see detail in \cite{cook2012pebbles}) of height $h$, where 
			$n_h = 2n_{h-1} + k^2 = (2^{h-1}-1)k^2 + 2^{h-1}$ 
		denotes the input size, and $n_1 = 1$.  
		Specifically, the input is a complete binary tree of height\footnote{Here the binary tree with a root and two leaves is deemed of height $2$.} $h$, in which every leaf is given an integer in $[k]$, and every internal node is given a matrix in $[k]^{k \times k}$.   One can naturally evaluate the binary tree in a bottom-up fashion and the output of the root node is defined as the output of $(\TEP,h)$.

	\item \emph{The BRS function}. Let $d\in \N$, $n=2^d$. The BRS function $\BRS_n: \{0,1\}^{2n} \times \{0,1\}^{2n} \to \{0,1\}$ is defined in \cite{BRS1993}    as follows. For every $x \in \{0,1\}^{2n}$, we write it as $x = (\ldots, x_{a1}, x_{a2}, \ldots)$, where $a\in \{0,1\}^d$ and $x_{a1}, x_{a2} \in \{0,1\}$. For every $y \in \{0,1\}^{2n}$, we write it as $y = (\ldots, x_{b1}, y_{b2}, \ldots)$, where $b\in \{0,1\}^d$ and $y_{b1}, y_{b2} \in \{0,1\}$. Define $\BRS_n(x,y) = 1$ if and only if $\sum_{a,b\in \{0,1\}^d} (-1)^{\inp{a}{b}} (x_{a1} + x_{a2}) (y_{b1} + y_{b2}) \equiv 0 \mod 3$, where $\inp{a}{b}$ is the usual inner product mod $2$, but the rest operations are performed mod $3$. More details of the definition is in \cite{BRS1993}. The BRS function is called as bilinear Sylvester function in \cite[Theorem 10.3.10]{wegener2000branching}.
	
	\item \emph{The indirect storage access function} $\ISA_n: [n] \times \{0,1\}^{n} \to \{0,1\}$ defined as follows, for $i\in [n]$ and $x \in \{0,1\}^n$, 
    $\ISA_n(i, x) = x_p$
    where $p$ is the integer represented by the length $\log n$  binary string $x_i x_{i+1} \cdots x_{i+\log n - 1}$ where the addition of indices is mod $n$.
	
	\item \emph{The iterated $\NAND_n$ function}. Define $\NAND_2: \{0,1\} \times \{0,1\} \to \{0,1\}$ by 
	    $\NAND_2(x,y)  = \lnot x \vee \lnot y$.
	For $n=2^h$, the iterated function $\NAND_n: \{0,1\}^n \to \{0,1\}$ is computed by the balanced read-once formula of height $h$ in which every gate is $\NAND_2$.
	
	\item \emph{The satisfiable Tseitin formulas}.
	Let $G=(V,E)$ be a graph, $|V| = n, |E|=m$, and 
        $c: V \to \{0,1\}$ 
    be a labelling function. The pair $(G,c)$ defines a Tseitin formula as a Boolean function on $\{0,1\}^E$, denoted by $\TS_{G,c}$, 
    \begin{equation}   \label{eq:Tseitin}
        \TS_{G,c}: \{0,1\}^E \to \{0,1\}, \quad
        x=(\cdots, x_e, \cdots) \mapsto  \bigwedge_{v\in V}  \left( \Big( \sum_{e \text{ is incident to } v} x_e \Big) \equiv c(v) \mod 2 \right).
    \end{equation}
    The Tseitin formula $\TS_{G,c}$ is said to be \emph{satisfiable} if the  Boolean function $\TS_{G,c}$ is not identically $0$. 
	
	\item \emph{The generalized G{\'a}l's function and Bollig-Wegener function}. 
    Let $G=(A\cup B, E)$ be a bipartite graph where $A$ and $B$ are the two parts of vertices. For $S \subseteq A$, let $\Nb(S) \subseteq B$ denote the set of neighbors of $S$ in $B$. The generalized G{\'a}l's  function on $G$, denoted by   
        $\GAL_G: \{0,1\}^{A} \to \{0,1\}$, 
    is defined as
                $\GAL_G(x) = 1$
            if and only if 
                $\Nb(x) = B$.
    The Bollig-Wegener function on $G$, denoted by 
            $\BW_G: \{0,1\}^{A} \times \{0,1\}^{B} \to \{0,1\}$,
    is defined as
                $\BW_G(x,y) = 1$
            if and only if      
                $\Nb(x) \cap y \neq \emptyset$.

	\item \emph{The GEN problem}.  For $m\ge 2$, let $n= \{(i,j): i, j \in [m-1], i \le j\}$, then $n=\frac{m(m-1)}{2}$. Every $X \in [m]^n$ defines an upper triangular matrix which can be thought of as a (commutative) multiplication table: $X_{ij} = k$ means $i*j = j*i= k$, where $1 \le i \le j \le m-1$ and $1 \le k \le m$.  Define $\GEN_n: [m]^n \to \{0,1\}$ as follows: $\GEN_n(X) = 1$ if and only if $m \in \langle 1 \rangle$, where $\langle 1 \rangle$ denotes  the set of elements generated starting from $1*1$ and using the  multiplication table $X$.  Note that if interpreted as a Boolean function, then $\GEN_n$ is defined on the domain $\{0,1\}^{n\log m} = \{0,1\}^{\Theta(n\log n)}$.

\end{itemize}

Lastly, for every Boolean function $f: \{0,1\}^n \to \{0,1\}$, let $\lnot f$ denote the negation of $f$, i.e., $(\lnot f)(x) = 1$ if and only if $f(x)=0$.

\section{The measures vs read-once branching program sizes}   \label{sec:property}

It is known that $\widehat{S}$ and $\widehat{C}$ are lower bounds for $\BPone$ and $\nBPone$, as proved in \cite{SS1993}\footnote{In fact, \cite{SS1993} proved  a stronger lower bound than $\widehat{S}$, but we focus on $\widehat{S}$ in this paper for its naturalness.} and \cite{BRS1993} respectively. The argument in \cite{SS1993}, \cite[Corollary 2]{BRS1993} in fact proved $\nBPone(f) \ge \widehat{C}(f,n/2)$, but it is easy to see that $\nBPone(f) \ge \widehat{C}(f)$ holds.

\begin{proposition}[\cite{SS1993,BRS1993}]   \label{prop:BP1-LB} 
Let $f: \{0,1\}^n \to \{0,1\}$. Then, $\BPone(f) \ge \widehat{S}(f)$,  $\nBPone(f) \ge \widehat{C}(f)$. 
\end{proposition}

In this section we will prove more lower bounds, study relations among measures and provide examples (many are derived from the existing literature) that separate the measures from each other and from read-once BP sizes in all possible cases. An interesting feature is that communication complexity often plays a role.

\subsection{$\widehat{P}$ is a lower bound for $\BP_1$}   \label{sec:weak-partition}

\begin{theorem}   \label{thm:hatP-BP1}
For every integer-valued function $f: \{0,1\}^n \to \N$,
    $\BP_1(f) \ge \widehat{P^+}(f)$.
When $f: \{0,1\}^n \to \{0,1\}$ is a Boolean function,
    $\BP_1(f) \ge \widehat{P}(f)$.
\end{theorem}

\begin{proof}
We adapt the proof strategy and ideas introduced in \cite{SS1993,BRS1993}. Let $B$ be a deterministic read-once BP computing $f$. We think of $B=(V(B), E(B))$ as a directed acyclic graph (DAG). Given a non-sink node $v \in V(B)$, let $x(v)$ denote the variable queried at node $v$. Given two nodes $u,v \in V(B)$, let 
    $u \le v$
denote the relation that either there is a directed path from $u$ to $v$ or $u=v$. Define
    \[
        X(u,v) = \{x(w): u \le w \le v\}.
    \]
Let $s, t_0, t_1 \in V(B)$ denote the source node and the two sink nodes (where $t_0$ and $t_1$ are the sink nodes with output $0$ and $1$, respectively). Note that 
    $X(s,t_0) = X(s,t_1)$. 
Without loss of generality we assume 
    $X(s,t_0) = X(s,t_1) = [n]$. 
Fix a parameter $1 \le k \le n$. Consider a mapping 
    \[
        \varphi_k: \{0,1\}^n \to E(B), \quad
            x \mapsto e=(u,v),
    \]
where $e=(u,v)$ is the  edge in the computation path of $x$ in $B$ that satisfies
    \[
        |X(s,u)| \le k < |X(s,v)|.
    \]
Observe that such edge $(u,v)$ is unique, hence $\varphi_k$ is well-defined. As usual,  we think of a subset $A \subseteq [n]$ as the corresponding subset of variables. Choose a subset $A_e \subseteq [n]$ such that 
    \[
        X(s,u) \subseteq A_e \subseteq X(s,v), \quad
        |A_e| = k.
    \]

Given two non-sink nodes $u,v \in V(B)$, let 
    $f_{u,v}: X(u,v) \backslash \{x(v)\} \to \{0,1\}$ 
denote the function computed by the sub-BP in $B$ with $u$ to be the source node, and 
    $f_{u,v}(y) = 1$
if and only if the input $y$ starts at $u$ and reaches $v$. Note that $f_{u,v}$ is well-defined because the BP $B$ is read-once. Given a non-sink node $v \in V(B)$, let $f_v: X(v,t_0) \cup X(v,t_1) \to \{0,1\}$ denote the function computed by the sub-BP in $B$  with $v$ to be the source node. The definitions of $f_{u,v}$ and $f_v$ can be extended to be defined on larger domains by ignoring the irrelevant variables. Let 
    $\mu(e) \in \{0,1\}$
denote the value marked on edge $e$, i.e., the edge $e=(u,v)$ tests whether $x(u) = \mu(e)$. Let $\Im(\varphi_k)$ denote the image of $\varphi_k$. With these notations, we define a function $f_e$ for every $e=(u,v) \in \Im(\varphi_k)$, 
    \begin{equation}  \label{eq:f_e}
        f_e: \{0,1\}^n \to \N, \quad
            x \mapsto f_{s,u}(x|_{A_e \backslash \{x(u)\}}) \cdot \big( x(u)==\mu(e) \big)  \cdot f_v(x|_{\overline{A_e}}).
    \end{equation}   
In other word, the function $f_e$ checks whether an input reaches the node $u$ and passes through the edge $e=(u,v)$ and is then computed by $f_v$. Hence, 
    \begin{equation}   \label{eq:f_e=f}
        f_e(x) =
        \begin{cases}
            f(x), &\quad x \in \varphi_k^{-1}(e), \\
            0, &\quad x \not\in \varphi_k^{-1}(e).
        \end{cases}
    \end{equation}
Equation \eqref{eq:f_e=f} implies that
    \begin{equation}   \label{eq:f-decompose}
        f(x) = \sum_{e \in \Im(\varphi_k)} f_e \quad
        \text{and} \quad
        \inp{f_e}{f_{e'}} = 0\ \forall\  e, e' \in \Im(\varphi_k), e\neq e'.
    \end{equation}
If we denote 
    $f_{s,u,v} (x|_{A_e}) = f_{s,u}(x|_{A_e - x(u)}) \cdot \big( x(u)==\mu(e) \big)$,
then,
    \[
        f_e = f_{s,u,v} \cdot f_v,
    \]
where $f_{s,u,v}$ 
is a Boolean-valued function defined on variables in $A_e$ and $f_v$ is an (non-negative) integer-valued function defined on variables in $\overline{A_e}$. By definition in Section \ref{sec:def-partition}, $f_e$ is an $k$-rectangle since $|A_e| = k$. Hence, by \eqref{eq:f-decompose} and Definition \ref{def:partition-numbers},
    \[
        |\Im(\varphi_k)| \ge \widehat{P^+}(f,k).
    \]
Since every node in $B$ has degree at most $2$, 
    $|V(B)| \ge |\Im(\varphi_k)| / 2$.
By observing that the edges in $\Im(\varphi_k)$ do not lie in the same path, the factor $2$ can be removed (see \cite[Theorem 2.4]{SS1993}), we omit this technical detail. As $1\le k \le n$ can be arbitrary,
    $\BP_1(f) \ge \max_k \widehat{P^+}(f,k) = \widehat{P^+}(f)$.
The lower bound for a Boolean function $f$ follows by noting that $\BP_1(f) = \BP_1(\lnot f)$.
\end{proof}

\begin{remark}
    The proofs for Proposition \ref{prop:BP1-LB}  in \cite{SS1993,BRS1993} both follow a similar strategy as shown above. Adopting the method in \cite{BRS1993} one could also generalize the measure $\widehat{P^+}(f)$ appropriately so that it becomes a lower bound for $\BP_k(f)$, see Section \ref{sec:read-k}.
\end{remark}

\subsection{$S^*$ (almost) characterizes $\OBDD$}   \label{sec:OBDD}

\begin{theorem}   \label{thm:OBDD-dS}
Let $f: \{0,1\}^n \to \{0,1\}$. Then,
    $S^*(f) \le \OBDD(f) \le 1+ n \cdot S^*(f)$.
\end{theorem}

\begin{proof}
    The claim is true if $f$ is a constant function, in which case $\OBDD(f) = S^*(f) = 1$. Assume now $f$ is not a constant function. 
    
    The lower bound:  By changing the names of the variables if necessary, assume that the size $\OBDD(f)$ is achieved with respect to the order of variables $x_1, \ldots, x_n$. Let $B_i = \{x_1, \ldots, x_i\}$ for $1\le i\le n$.  For $A\subseteq \{x_1, \ldots, x_n\}$, let $R(A) = \text{the set of distinct rows of the matrix } f_A$, i.e., it is the set of all distinct subfunctions $f_{A,\alpha}$ where $\alpha \in \{0,1\}^A$.  Consider $R(B_i)$. Let 
$U(B_i) = \{g \in R(B_i) : g|_{x_{i+1}=0} \neq g|_{x_{i+1}=1}\}$, i.e., it is the subset of subfunctions in $R(B_i)$ that essentially depend on $x_{i+1}$. Let $W(B_i) = R(B_i) \backslash U(B_i)$, i.e., the subset of subfunctions in $R(B_i)$ that do not essentially depend on $x_{i+1}$. Let $u(B_i) = |U(B_i)|$ and $w(B_i) = |W(B_i)|$. So, 
\begin{equation}   \label{eq:u-plus-w}
S(f_{B_i})=|R(B_i)| = u(B_i) + w(B_i).
\end{equation}
By  \cite[Theorem 3.1.4]{wegener2000branching},
    \begin{equation}  \label{eq:OBDD-wegener}
        \OBDD(f) = 1+ \sum_{i=1}^n u(B_i).
    \end{equation}
We claim $w(B_i) \le S(f_{B_{i+1}})$. Since we assume $f$ is not a constant function,  $S(f_{B_{n}}) = u(B_n) = 2$. This and \eqref{eq:u-plus-w} together imply  that $S(f_{B_i}) \le \sum_{j=i}^n u(B_j) \le \OBDD(f)$ holds for every $1\le i \le n$. Hence, 
    $S^*(f) = \min_\sigma \max_{1 \le i \le n} S(f_{\sigma([i])})
    \le \max_{1 \le i \le n} S(f_{B_i})
    \le \OBDD(f)$.

We proceed to show the claim. Consider a mapping 
\[
\phi: W(B_i) \to R(B_{i+1}),
\quad
g \mapsto \phi(g)
\]
where $\phi(g) = g|_{x_{i+1}=0}$. It suffices to show $\phi$ is injective.  Indeed, if $g,g' \in W(B_i)$ and $g \neq g'$, then there exists $(x_{i+1}, x_{i+2}, \ldots, x_n)$ such that $g(x_{i+1}, x_{i+2}, \ldots, x_n) \neq g'(x_{i+1}, x_{i+2}, \ldots, x_n)$. Since $g,g' \in W(B_i)$, one has 
\begin{align*}
\phi(g)(x_{i+2}, \ldots, x_n) 
= g(0, x_{i+2}, \ldots, x_n) 
&= g(x_{i+1}, x_{i+2}, \ldots, x_n) \\
&\neq g'(x_{i+1}, x_{i+2}, \ldots, x_n) \\
&= g'(0, x_{i+2}, \ldots, x_n) = \phi(g')(x_{i+2}, \ldots, x_n). 
\end{align*}
That is, $\phi(g) \neq \phi(g')$ as desired.

    The upper bound: suppose $S^*(f)$ is minimized with respect to the order of variables $x_1, \ldots, x_n$, denote this order by $\pi$. Let $\pi\text{-}\OBDD(f)$ denote the least OBDD size of $f$ when variables are queried with respect to the order $\pi$.      Use the notation in the lower bound proof, by \eqref{eq:u-plus-w} and \eqref{eq:OBDD-wegener},
        \[
            \OBDD(f) \le \pi\text{-}\OBDD(f) 
            = 1 + \sum_{i=1}^n S(f_{B_i}) 
            \le 1 + n \max_{1 \le i \le n} S(f_{B_i}) 
            = 1 + n \max_{1 \le i \le n} S(f_{\pi([i])})
            = 1 + n\cdot S^*(f),
        \]
    where the last step follows by the assumption that $S^*(f)$ is minimized at $\pi$.
\end{proof}

\begin{remark}
    The measure $S^*(f)$ as a lower bound for $\OBDD(f)$ has essentially been applied in the literature,  e.g., \cite[Proposition 1]{razgon2014obdds} as well as in \cite{wegener2000branching} etc, however, to the best knowledge of the authors there was no formal proof for this fact. 
\end{remark}

\begin{corollary}  \label{cor:S-OBDD}
    $\OBDD(f) \ge S(f)$.
\end{corollary}

\subsection{Relations and separations}  \label{sec:properties}

In this section we focus exclusively on Boolean functions $f: \{0,1\}^n \to \{0,1\}$.  

\begin{theorem}   \label{thm:relation}
Let $f: \{0,1\}^n \to \{0,1\}$ be a Boolean function. 
\begin{enumerate}[(1)]
    \item $\widehat{S}(f) \le S(f)$, $\widehat{C}(f) \le C(f)$, 
        $\widehat{P}(f) \le P(f)$, $\widehat{C}(f) \le \widehat{P}(f)$, 
        $C(f) \le P(f)$,
    \item $P(f)/2 \le S(f) \le 2^{C(f)}$,
    \item $S(f) = O(2^n/n)$, $P(f) \le 2^{n/2+1} = O(2^{n/2})$.
    \item for every $\delta > 1$, most $f$ satisfies $\widehat{S}(f) \ge 2^n/n^\delta$,
    \item most $f$ satisfies $\widehat{C}(f) \ge \Omega(2^{n/2} / \log n)$.
\end{enumerate}
\end{theorem}

\begin{proof}
(1). These follow directly from definitions. 

(2). It suffices to show that for every subset $A \subseteq [n]$, 
    $P(f_A)/2 \le S(f_A) \le 2^{C(f_A)}$. 
The first inequality follows from 
    $P^{+}(f_A) \le S(f_A)$
and 
    $P^{+}(\lnot f_A) \le S(\lnot f_A)$
and noting that 
    $S(f_A) = S(\lnot f_A)$.
For the second, by the definition of $C(f, A)$, suppose $C(f, A) = r$, then
    \[
        f_A = g_1 \lor \cdots \lor g_r,
    \]
where each matrix $g_i$ has rank $1$. For every $1\le i \le r$, let $X_i$ denote the unique nonzero row in the matrix $g_i$. Then, the above equation implies that each row of the matrix $f_A$ can be written in the form $\lor_{i \in S} X_i$ for some subset $S \subseteq [r]$. Hence, $f_A$ contains at most $2^r  = 2^{C(f, A)}$ distinct rows. 

(3). By definition, 
    \[
        S(f) = \max_{1\le k\le n} \min_{A \subseteq [n], |A| = k} \nrows(f_A) \le \max_{1\le k\le n} \min\{2^k, 2^{2^{n-k}}\} = O(2^n/n). 
    \]
By \eqref{eq:CC-ub} and \eqref{eq:P-and-CC}, 
    $P(f) \le 2^{\CC(f)} \le 2^{n/2+1}$.
    
(4). Observe  that for a random Boolean matrix $M_{a\times b} \in  \{0,1\}^{a \times b}$ of $a$ rows and $b$ columns, $\Pr[M \text{ contains at least } 2 \text{ identical rows}] \le \frac{a^2}{2^b}$. Now consider $\widehat{S}(f)$. Consider a random function $f: \{0,1\}^n \to \{0,1\}$. Then, for every $A \subseteq [n]$ of size $|A| = k$, the matrix $f_A$ is a random Boolean matrix of $a=2^k$ rows and $b=2^{n-k}$ columns. Since there are ${n \choose k}$ ways of choosing subsets $A \subseteq [n]$, 
    \[
        \Pr[\text{There exists } A \text{ of size } |A| = k, \text{ s.t. } f_A \text{ contains at least } 2 \text{ identical rows}] \le {n \choose k} \cdot \frac{2^{2k}}{2^{2^{n-k}}}.
    \]
Let $\delta > 1$. Choose $k=n- \delta \log n$. Then, the above probability is upper bounded by $2^{(2+\delta)n}/2^{n^\delta}$. Hence, 
    \begin{align*}
        &\Pr[\max_{|A| = n- \delta \log n} \mult(f_A) = 1] \\
        &= \Pr[\text{For every } A \text{ of size } |A|=n- \delta \log n, \text{ rows in } f_A \text{ are all distinct}] \\
        &\ge 1 - \frac{2^{(2+\delta)n}}{2^{n^\delta}}.
    \end{align*}
Equivalently, for a random function $f$, with probability at least $1- 2^{(2+\delta)n}/2^{n^\delta}$, $\widehat{S}(f) \ge 2^{n-\delta \log n}$.

(5). Let $r = \widehat{C}(f)$, then $\widehat{C}(f,n/2) \le r$ (if $n$ is odd one chooses $(n+1)/2$). By definition of $\widehat{C}(f,n/2)$, 
\[
f = f_1 \lor \cdots \lor f_r
\]
where each $f_i$ is an $n/2$-rectangle. That is, $f_i = f_{i1} \land f_{i2}$ where $f_{i1}: \{0,1\}^{n/2} \to \{0,1\}$ and $f_{i2}: \{0,1\}^{n/2} \to \{0,1\}$ are two Boolean functions each is defined on $n/2$ bits. This shows the formula size of $f$ is at most $O(r \cdot 2^{n/2})$. It is well known that most Boolean functions have formula size $\Omega(2^n/\log n)$ (see, e.g., \cite{wegener1987complexity}). Hence, $r \ge \Omega(2^{n/2}/\log n)$. 
\end{proof}

\begin{corollary}   \label{cor:bCC-to-size-and-cover}
Let $f: \{0,1\}^n \to \{0,1\}$ be a Boolean function. Then,
\[
\log \log S(f) \le \log C(f) = \NCC(f) \le \log P(f) \le \CC(f) \le  1+ \log S(f). 
\]
\end{corollary}

\begin{proof}
Inequalities except the last one follow from Theorem \ref{thm:relation} and \eqref{eq:P-and-CC}. The last inequality can be easily proved by the definition of communication complexity (see \cite{kushilevitz1997communication}). We omit the detail. 
\end{proof}

We now show the inequality $C(f) \ge \log S(f)$ is tight up to a polynomial (the rest inequalities in Theorem \ref{thm:relation} are easily seen to be tight).

Viewing $\SEQ_n$ as a Boolean function on $2n+\log n$ bits, the max-min deterministic and nondeterministic communication complexity of $\SEQ_n$ are given below. 
\begin{proposition} \label{prop:SEQ}
$\CC(\SEQ_n) = \CC(\lnot \SEQ_n) = \Theta(n)$, $\NCC(\SEQ_n) = \Theta(n)$, $\NCC(\lnot \SEQ_n) \le 1+2\log n = O(\log n)$. 
\end{proposition}

\begin{proof}
\cite[Example 7.9]{kushilevitz1997communication} shows that $\min_{A \subseteq [n], |A| = n+\frac{\log n}{2}} \ccmat((\SEQ_n)_A) = \Theta(n)$, hence $\CC(\SEQ_n) = \Theta(n)$. 
The proof for \cite[Example 7.9]{kushilevitz1997communication} uses the fact that $\ccmat(\EQ_n) \ge n$. Since $\nccmat(\EQ_n) \ge n$ also holds (see \cite{kushilevitz1997communication}), it is not hard to see that the  proof for \cite[Example 7.9]{kushilevitz1997communication} can be adapted to show $\NCC(\SEQ_n) = \Theta(n)$. We omit the detail. Since $\nccmat(\lnot \EQ_n) \le 1+ \log n$ (see \cite{kushilevitz1997communication}),  then $\NCC(\lnot \SEQ_n) \le 1+ 2 \log n$.
\end{proof}

In viewing of Proposition \ref{prop:BP1-LB} and Theorem \ref{thm:hatP-BP1}, it is natural to ask whether $S$, $P$ and $C$ are also lower bounds for $\BPone$ or $\nBPone$. Perhaps not surprisingly, none of them is.

\begin{proposition}   \label{prop:sep-SEQ}
 $\BPone(\SEQ_n) = O(n^2)$,  $S(\SEQ_n) = 2^{\Theta(n)}$, 
 $P(\SEQ_n) \ge C(\SEQ_n) = 2^{\Theta(n)}$.
\end{proposition}

\begin{proof}
Consider a BP that firstly reads the index $i$, using $O(2^{\log n}) = O(n)$ number of nodes, then for each index $i$, it continues to compute an equality function using $O(n)$ nodes. This BP is read-once. Hence, $\BPone(\SEQ_n) = O(n^2)$. 
 Lower bounds for $S$ and $C$ follow from Corollary \ref{cor:bCC-to-size-and-cover} and Proposition \ref{prop:SEQ}.
\end{proof}

\begin{proposition}   \label{prop:sep-SIEQ} 
$S(\lnot \SEQ_n) = 2^{\Theta(n)}$, 
$P(\lnot \SEQ_n) = 2^{\Omega(\sqrt{n})}$,
$C(\lnot \SEQ_n) = O(n^2)$.
\end{proposition}

\begin{proof}
Apply Corollary \ref{cor:bCC-to-size-and-cover} and Proposition \ref{prop:SEQ} and Equation \eqref{eq:P-and-CC} and \eqref{eq:NCC-C}.
\end{proof}

Theorem \ref{thm:relation} shows that $\widehat{S}(f)$, $\widehat{P}(f)$ and $\widehat{C}(f)$ are good lower bounds for $\BP_1(f)$ and $\nBP_1(f)$, respectively, in the following sense: since $\BP_1(f) \le O(2^n/n)$ and $\nBP_1(f) \le O(2^{n/2})$ (see, e.g., \cite[Lemma 3.6]{NeciBoundTight2016}), property (4) and (5)  of Theorem \ref{thm:relation} show that $\widehat{S}$, $\widehat{P}$ and $\widehat{C}$ provide almost tight exponential lower bounds for $\BP_1(f)$ and $\nBP_1(f)$, respectively, 
for \emph{most} functions. Similarly, $S(f)$ is a good lower bound for $\OBDD(f)$. Nonetheless, they do not characterize the corresponding BP sizes.  For example, $\BPone(\PARITY_n) = \Theta(n)$, but $S(\PARITY_n) = 2$ which implies that $\widehat{S}(\PARITY_n)$, $\widehat{P}(\PARITY_n)$ and $\widehat{C}(\PARITY_n)$  all equal to $2$ via Theorem \ref{thm:relation}. In fact, the gaps can be exponential.

\begin{proposition}[\cite{SS1993,BRS1993}]   \label{prop:sep-hat-C-from-hat-s}
$\widehat{C}(\clique_{n,n/2}) = 2^{\Theta(n)}$,  $\widehat{S}(\clique_{n,n/2}) \le 2^{49} = O(1)$.
\end{proposition}

\begin{proof}
The lower bound for $\widehat{C}$ is shown in  \cite[Theorem 3]{BRS1993}. \cite{SS1993} mentioned that ``the uniform weighting provably cannot work'', using our notation, this is equivalent to say that $\widehat{S}(\clique_{n,n/2}) = n^{O(1)}$. We proceed to give an explicit bound.  

Let $N = n(n-1)/2$. For every $1 \le k \le N$, we choose a subset $A \subseteq [N]$ so that $\mult((\clique_{n,n/2})_{A})$ is large, as follows. Let $p_k$ be the largest integer so that $T_k = p_k(p_k-1)/2 \le k$. Choose $p_k$ vertices from $n$ vertices, and choose all the $T_k$ edges on these $p_k$ vertices to be in $A$, and choose the remaining $k-T_k$ edges, if any, arbitrarily. By our choice, $k-T_k \le p_k \le 3\sqrt{k}$. 

Let $(\alpha, \beta) \in \{0,1\}^{A} \times \{0,1\}^{\overline{A}} = \{0,1\}^N$, let $\alpha_{p_k}$ denote the subgraph on the $p_k$ vertices given by $\alpha$. We say the subgraph $\alpha_{p_k}$ is a $q$-clique if it contains exactly a clique of size $q$ and isolated $p_k-q$ vertices, where $0 \le q \le p_k$. Observe that  $\clique_{n,n/2}(\alpha,\beta) = 1$ implies that $\alpha_{p_k}$ must be a $q$-clique for some $0 \le q \le p_k$. Alternatively, if for every $0 \le q \le p_k$, the subgraph $\alpha_{p_k}$ is not a $q$-clique, then the subfunction $(\clique_{n,n/2})_{A,\alpha} = 0$. Since the number of $q$-cliques on $p_k$ vertices is ${p_k \choose q}$, we get the number of  choices of $\alpha$ such that $(\clique_{n,n/2})_{A,\alpha} = 0$  is at least
\[
\left(2^{T_k} - \sum_{q=0}^{p_k} {p_k \choose q} \right) \cdot 2^{k-T_k}
=  (2^{T_k} - 2^{p_k})  \cdot 2^{k-T_k}
\ge 2^k - 2^{6\sqrt{k}}.
\]
This shows $\mult((\clique_{n,n/2})_{A}) \ge  2^k - 2^{6\sqrt{k}}$. Hence, $2^k/ \mult((\clique_{n,n/2})_{A}) \le 2$ when $k \ge 49$. 
\end{proof}

The proof for $\widehat{S}(\clique_{n,n/2}) = O(1)$ in Proposition \ref{prop:sep-hat-C-from-hat-s} lies in the fact that there are many constant $0$ subfunctions in $\clique_{n,n/2}$. This may seem too special. In Section \ref{sec:TEP} we provide another separation example of $\BPone$ from $\widehat{S}$ via the function TEP. Note that obviously $\widehat{S}(f)=\widehat{S}(\lnot f)$ and $\widehat{P}(f)=\widehat{P}(\lnot f)$ for every Boolean function $f$, but this is not true for $\widehat{C}$.

\begin{proposition}  \label{prop:negation-clique}
$\widehat{C}(\lnot \clique_{n,n/2})  \le \nBPone(\lnot \clique_{n,n/2}) = O(n^4)$, 
$\NCC(\lnot \clique_{n,n/2}) = O(\log n)$,  $\widehat{P}(\lnot \clique_{n,n/2}) = 2^{\Theta(n)}$.
\end{proposition}

\begin{proof}
The first upper bound follows from Proposition \ref{prop:BP1-LB} and $\nBPone(\lnot \clique_{n,n/2}) = O(n^4)$ showed in 
\cite[Theorem 5]{BRS1993}.

For the second upper bound, for any subset $A \subseteq [n(n-1)/2]$, consider $\nccmat(\lnot \clique_{n,n/2})_A$. By the proof of \cite[Theorem 5]{BRS1993}, a graph $G$ is not an exact $\ceil{n/2}$-clique if and only if at least one of the following is true:
\begin{itemize}
    \item[(a)] there are two edges $(v_1, v_2)$ and $(v_3, v_4)$ in $G$ such that $v_1 \neq v_3$ and $(v_1, v_3)$ is not an edge;
    \item[(b)] there exists at least one vertex $v$ whose degree differs from both $0$ and $\ceil{n/2}$;
    \item[(c)] $G$ is empty.
\end{itemize}
Alice and Bob can verify whether $G$ is empty using $O(1)$ communication. For case (a), the prover can give the four vertices $v_1, v_2, v_3, v_4$ as a proof, which has length $O(\log n)$, and Alice and Bob need to use only $O(1)$ communication to verify it. For case (b), the prover can give the name of $v$ as a proof, which has length $O(\log n)$, and Alice and Bob need to use $O(\log n)$ communication to verify it. Hence,
    $\nccmat(\lnot \clique_{n,n/2})_A = O(\log n)$, 
independent of the choice of $A$. Hence,
    $\NCC(\lnot \clique_{n,n/2}) = O(\log n)$.

Lastly, 
$\widehat{P}(\lnot \clique_{n,n/2}) = \widehat{P}(\clique_{n,n/2}) \ge  \widehat{C}(\clique_{n,n/2}) = 2^{\Theta(n)}$.
\end{proof}

\begin{proposition} \label{prop:sep-hat-s-from-C}
$\widehat{S}(\pi_n) \ge 2^{\sqrt{n/\log n}-1}$, $P(\pi_n) \le 16 n$, $\CC(\pi_n) \le \log n + 4$, $\nBP_1(\pi_n) = O(n)$.
\end{proposition}

\begin{proof}
\cite[Theorem 16.8]{jukna2012boolean} showed that $\pi_n$ is $(\sqrt{n/\log n}-1)$-mixed, hence $\widehat{S}(\pi_n) \ge 2^{\sqrt{n/\log n}-1}$. \cite[Theorem 16.8]{jukna2012boolean} also showed $\nBP_1(\pi_n) = O(n)$. For the rest two, \eqref{eq:P-and-CC} implies it suffices to show $\CC(\pi_n) \le \log n + 4$. Indeed, for every $1 \le k \le n$, choose $A = [k]$. It is easy to see that
    $\ccmat((\pi_n)_A) \le \log n + 4$.
\end{proof}

Lastly, we consider separation of $\widehat{C}$ from $\nBPone$.

\begin{proposition} \label{prop:small-S-large-NBP}
The following hold. 
\begin{enumerate}[(1)]
    \item There exists a Boolean function $f: \{0,1\}^n \to \{0,1\}$ that depends on all its input variables,  such that $\nBP_k(f) = 2^{\Omega(\frac{n}{4^k k^3})}$ and $S(f) \le 16$. 
    \item There exists a Boolean function $f: \{0,1\}^n \to \{0,1\}$ that depends on all its input variables,  such that the circuit size of $f$ is $\Omega(2^{n/2}/n)$ and $S(f) \le 16$. 
\end{enumerate}
\end{proposition}

\begin{proof}
(1) Assume $n$ is even (it will be clear that odd $n$ can be handled similarly). Write $z \in \{0,1\}^n$ as $z=(x,y)$ where $x,y \in \{0,1\}^{n/2}$. Define $f(z) = g(x) \land  \PARITY_{n/2}(y)$. Trivially, $f$ depends on all its variables as long as $g$ depends on every variable in $x$. Observe that $\nBP_k(f) \ge \nBP_k(g)$. On the other hand, it is easy to see that $\CC(f) \le 2$, hence $S(f) \le 16$ by Corollary \ref{cor:bCC-to-size-and-cover} . Set $g = \BRS_{n/2}$ gives the result. 

(2) This can be proved similarly as (1).
\end{proof}

\begin{figure}[h!]        
\centering
        \includegraphics[scale=0.9]{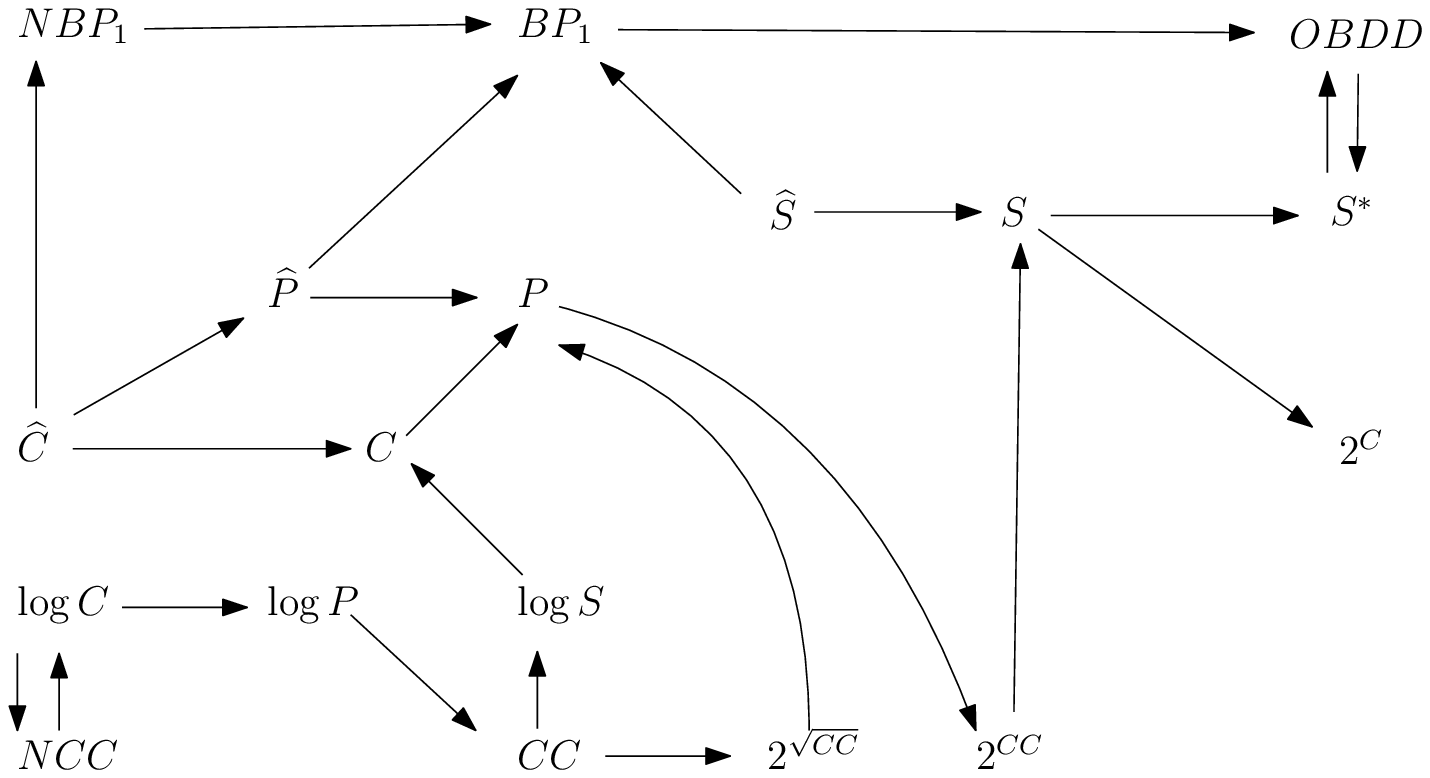}
       \caption{The relation among measures for Boolean functions $f: \{0,1\}^n \to \{0,1\}$. A directed edge from $a$ to $b$ indicates $a \lesssim b$.}    \label{fig:relation}
\end{figure}

We summarize the relations and separations  in Figure \ref{fig:relation}  and Table \ref{table:compare}, respectively. Note that all measures in Table \ref{table:compare} are lower bounds for $\OBDD(f)$. The measures $\OBDD(f)$ and $2^{C(f)}$  in Figure \ref{fig:relation} can be shown to be incomparable by the method in Proposition \ref{prop:small-S-large-NBP}.

\begin{table}[h!]
\centering
\caption{Comparison of measures. The notation $\ll$ (resp.  $\gg$) indicates an exponential separation in the direction Row $\ll$ (resp.  $\gg$)  Column. All measures in the table are lower bounds for $\OBDD$.}
\label{table:compare}
\resizebox{\textwidth}{!}{
\begin{tabular}{|c|c |c |c |c|c|c|c|c|} 
\hline
                            		& 
           $\nBP_1$ 		& 
           $\BP_1$ 		& 
           $\widehat{C}$ 	& 
           $C$ 			&
           $\widehat{P}$ 	& 
           $P$ 			&
           $\widehat{S}$ 	& 
           $S$ 			\\
\hline
		\multirow{2}{*}{$\nBP_1$} 						&
		\multirow{2}{*}{$=$}    								&
		$\le$												&
		$\ge$: \cite[Thm 1]{BRS1993}					&
		$\ll$: Prop \ref{prop:sep-SEQ}					&
		$\ll$: Prop \ref{prop:negation-clique}			   &
		$\ll$: Prop \ref{prop:sep-SEQ}   &
   		$\ll$: Prop \ref{prop:sep-hat-s-from-C}			&
   		$\ll$: Prop \ref{prop:sep-SEQ}					\\  
        														&    
   															&
   		$\ll$:  \cite[Thm 16.8]{jukna2012boolean} 		&
   		$\gg$: Prop \ref{prop:small-S-large-NBP}		&
   		$\gg$: Prop \ref{prop:small-S-large-NBP}		&
   		$\gg$: Prop \ref{prop:small-S-large-NBP}   &
		$\gg$: Prop \ref{prop:small-S-large-NBP}   &
   		$\gg$: Prop \ref{prop:small-S-large-NBP}		&
   		$\gg$: Prop \ref{prop:small-S-large-NBP}		\\   
\hline
   		\multirow{2}{*}{$\BP_1$} 							&
   		\multirow{2}{*}{} 									&
   		\multirow{2}{*}{$=$} 								&
   		$\ge$: \cite[Thm 1]{BRS1993}					&
   		$\ll$: Prop \ref{prop:sep-SEQ}					&
   		$\ge$: Thm \ref{thm:hatP-BP1}   &
		$\ll$: Prop \ref{prop:sep-SEQ}   &
   		$\ge$:	\cite[Thm 2.1]{SS1993}					&
   		$\ll$: Prop \ref{prop:sep-SEQ}					\\  
        														&    
															&
  															&
   		$\gg$: Prop \ref{prop:small-S-large-NBP}		&
   		$\gg$: Prop \ref{prop:small-S-large-NBP}		&
   		$\gg$: Prop \ref{prop:sep-hat-s-from-C}   &
		$\gg$: Prop \ref{prop:sep-hat-s-from-C}   &
   		$\gg$: Prop \ref{prop:sep-hat-C-from-hat-s}		&
   		$\gg$: Prop \ref{prop:small-S-large-NBP}		\\   
\hline
   		\multirow{2}{*}{$\widehat{C}$} 					&
    		\multirow{2}{*}{} 									&
        	\multirow{2}{*}{} 									&
   		\multirow{2}{*}{$=$} 								&
   		$\le$: Thm \ref{thm:relation}						&
   		$\le$: Thm \ref{thm:relation}   &
		$\le$: Thm \ref{thm:relation}   &
   		$\ll$: Prop	\ref{prop:sep-hat-s-from-C}			&
   		$\le$: Thm \ref{thm:relation}						\\  
        														&    
   															&
   															&
   															&
   		$\ll$: Prop \ref{prop:sep-SEQ}					&
   		$\ll$: Prop \ref{prop:negation-clique}			  &
		$\ll$: Prop \ref{prop:sep-SEQ}   &
   		$\gg$:	Prop \ref{prop:sep-hat-C-from-hat-s}		&
    		$\ll$: Prop \ref{prop:sep-SEQ}					\\   
\hline
   		\multirow{2}{*}{$C$} 								&
    		\multirow{2}{*}{} 									&
        	\multirow{2}{*}{} 									&
         	\multirow{2}{*}{} 									&
   		\multirow{2}{*}{$=$} 								& 
   		$\ll$: Prop \ref{prop:negation-clique}  			&
		$\le$: Thm \ref{thm:relation}                  &
   		$\ll$: Prop \ref{prop:sep-hat-s-from-C}			&
   		$\le$: Thm \ref{thm:relation}						\\  
        														&    
															&
															&
   															&
 															&
 		$\gg$: Prop \ref{prop:sep-SEQ}   &
		$\ll$: Prop \ref{prop:sep-SIEQ}   &										
   		$\gg$:	Prop \ref{prop:sep-hat-C-from-hat-s}		&
    		$\ll$: Prop \ref{prop:sep-SIEQ} 			\\   
\hline
     		\multirow{2}{*}{$\widehat{P}$} 					&
    		\multirow{2}{*}{} 									&
        	\multirow{2}{*}{} 									&
         	\multirow{2}{*}{} 									&
         	\multirow{2}{*}{} 									&
        	\multirow{2}{*}{$=$}                               &
		    $\le$: Thm \ref{thm:relation}   &
       		$\ll$: Prop \ref{prop:sep-hat-s-from-C} 			& 
   		    $\le$: Thm \ref{thm:relation}						\\  
        														&    
   															&
   															&
 															&
   															&
   		                                                        &
		    $\ll$: Prop \ref{prop:sep-SEQ}                      &
  		    $\gg$: Prop \ref{prop:sep-hat-C-from-hat-s}	        &
    		$\ll$: Prop \ref{prop:sep-SEQ}          		\\   
\hline
     		\multirow{2}{*}{$P$} 					&
    		\multirow{2}{*}{} 									&
        	\multirow{2}{*}{} 									&
         	\multirow{2}{*}{} 									&
         	\multirow{2}{*}{} 									&
    	    \multirow{2}{*}{}                                  &
    		\multirow{2}{*}{$=$}                               &
   		    $\ll$: Prop \ref{prop:sep-hat-s-from-C}             & 
   		    $\le$: Thm \ref{thm:relation}						\\  
        														&    
   															&
   															&
 															&
   															&
   		                                                    &
		                                                    &
  			$\gg$: Prop \ref{prop:sep-hat-C-from-hat-s}	    &
    		$\ll$: Prop \ref{prop:sep-hat-s-from-C} 		\\   
\hline
     		\multirow{2}{*}{$\widehat{S}$} 					&
    		\multirow{2}{*}{} 									&
        	\multirow{2}{*}{} 									&
         	\multirow{2}{*}{} 									&
         	\multirow{2}{*}{} 									&
    	                                                       &
	                                                           &
   		\multirow{2}{*}{$=$} 								& 
   		$\le$: Thm \ref{thm:relation}						\\  
        														&    
   															&
   															&
 															&
   															&
       		                                               &
		                                                      &
  															&
    		$\ll$: Prop \ref{prop:sep-hat-C-from-hat-s}		\\   
\hline
\end{tabular}
}  
\end{table}

\begin{remark}
Proposition \ref{prop:sep-SEQ} and Proposition \ref{prop:small-S-large-NBP} show that there are exponential separations between circuit size (or branching program size, or formula size) of $f$ and $S(f)$ in both directions. In fact, simple functions can provide separations such as $\SEQ \in \AC^0$. As mentioned in the introduction of \cite{SS1993}, Uhlig  showed that if the average number of subfunctions is a constant, then the circuit size is linear. The proof for (2) in Proposition \ref{prop:small-S-large-NBP} shows that this is not true for $S(f)$, $S(f)$ can be exponentially smaller than the average number of subfunctions. 
\end{remark}

\subsection{Lower bound measures for read-$k$ BPs}   \label{sec:read-k}

Although we will not elaborate on these,  in this subsection we define lower bound measures for $\BP_k$ and $\nBP_k$. 
    We say $d$  (not necessarily disjoint) subsets 
    $A_1, \ldots, A_d \subseteq [n]$ 
    is a \emph{read-$k$ partition of size $\ell$} for $[n]$, if they satisfy the following conditions: 
    (1) $\cup_{i=1}^d A_i = [n]$;
    (2) $|A_i| =  \ell$ for every $i\le d-1$, and 
        $|A_d| \le \ell$; and
    (3) every element $t \in [n]$ appears in at most $k$ $A_i$'s.
    A Boolean function 
        $g: \{0,1\}^n \to \{0,1\}$ 
    is said to be a \emph{$(k,\ell)$-hyperrectangle}\footnote{This is called $(k,n/\ell)$-rectangle in \cite{wegener2000branching}. If we view $d$ as the dimension of a hyperrectangle, we remark that a Boolean function $g$ can be viewed as a $(k,\ell)$-hyperrectangle in different dimensions. For example, consider $(2,n/4)$-hyperrectangles. One can have the following two distinct read-$2$ partitions of size $n/4$ for $[n]$ as follows: 
    (1) an equi-partition into four disjoint subsets, so $d=4$; (2) equi-partition of $[n]$ into $8$ disjoint subsets, say $B_1, \ldots, B_8$, then set $A_i = B_i \cup B_{i+1}$ for $i=1,\ldots, 7$, and set $A_8 = B_8 \cup B_1$, so $d=8$. Though $(k,\ell)$-hyperrectangles can be of different dimensions, the dimension satisfies $d \le kn/\ell$. } 
    if there exists a read-$k$ partition of size $\ell$ for $[n]$, say 
        $(A_1, \ldots, A_d)$,
    such that for every $x \in \{0,1\}^n$,
        \begin{equation}   \label{eq:def-k-ell-hyper}
            g(x) = \bigwedge_{i=1}^d g_i(x_i), \quad 
            \text{where } x_i = x|_{A_i},\ 
            g_i: \{0,1\}^{A_i} \to \{0,1\}
            \text{ is a Boolean function}.
        \end{equation}
Let $f: \{0,1\}^n \to \{0,1\}$ be a Boolean function. We define $\widehat{\cC_k}(f,\ell)$ to be the minimal integer $r$ such that
    $f = \lor_{j=1}^r f_j$
where each $f_j: \{0,1\}^n \to \{0,1\}$ is a $(k,\ell)$-hyperrectangle. Finally, define\footnote{Here the  exponent $\ell/kn$ in defining $\widehat{\cC_k}$ is the inverse of the maximal possible dimension of the $(k,\ell)$-hyperrectangles, see the previous footnote.} 
    $\widehat{\cC_k}(f) = \left( \max_{1 \le \ell \le n} \widehat{\cC_k}(f,\ell)^\ell \right)^{\frac{1}{kn}}$.

We proceed to define the counterpart for deterministic case. 
An integer-valued function 
    $g: \{0,1\}^n \to \N$ 
is said to be a \emph{$(k,\ell)$-hyperrectangle} if a similar condition as \eqref{eq:def-k-ell-hyper} holds, except that the $\land$ is replaced by the integer multiplication, and the last function $g_d: \{0,1\}^{A_d} \to \N$ is integer-valued. Given 
    $f: \{0,1\}^n \to \N$ 
to be an integer-valued function. Similar to Definition \ref{def:partition-numbers}, define 
    $\widehat{\cP^+_k}(f,\ell)$
to be the minimal integer $r$ such that $f$ can be decomposed into $r$ orthogonal $(k,\ell)$-hyperrectangles. Then, we define
    $\widehat{\cP^+_k}(f) = \left( \max_{1 \le \ell \le n} \widehat{\cP^+_k}(f,\ell)^\ell \right)^{\frac{1}{kn}}$.
When 
    $f: \{0,1\}^n \to \{0,1\}$
is a Boolean function, we define
    $\widehat{\cP_k}(f) = \max \{\widehat{\cP^+_k}(f), \widehat{\cP^+_k}(\lnot f)\}$.

\begin{proposition}  \label{prop:read-k-lb}
    Let 
        $f: \{0,1\}^n \to \{0,1\}$
    be a Boolean function. Then, 
    \[
        \BP_k(f) \ge \widehat{\cP_k}(f), 
        \quad
        \nBP_k(f) \ge \frac{1}{2} \sqrt{\widehat{\cC_k}(f)}.
    \]
    If 
        $f: \{0,1\}^n \to \N$
    is integer-valued, then
        $\BP_k(f) \ge \widehat{\cP^+_k}(f)$.
\end{proposition}

The lower bound for $\nBP_k$ by $\widehat{\cC_k}(f)$ is proved in \cite{BRS1993} (see also \cite{wegener2000branching}), and it is shown that 
    $\widehat{\cC_k}(\BRS_n) = 2^{\Omega(\frac{n}{k^3 4^k})}$. 
The lower bounds for $\BP_k$ can be proved by a small modification of the proof for Theorem \ref{thm:hatP-BP1}, using the method in \cite{BRS1993}. We omit the detail.

\section{$\widehat{S}$ is small for TEP}   \label{sec:TEP}

Iwama et al. \cite{iwama2018read} showed that $\BPone(\TEP,h) \ge k^h$ as long as $h\le (\frac{1}{3} - \frac{1}{2\log k})k$.
The approach in \cite{iwama2018read} is to directly exploit properties satisfied by a read-once BP for $(\TEP,h)$, and seems hard to be generalized to the read-$k$ case. It would be desirable to give an alternative proof that is amenable for a possible generalization to read-$k$. Unfortunately, below we show that  $\widehat{S}$ does not work: $\widehat{S}(\TEP,h)$ is small. In  this section we will use the definition of  $\widehat{S}$ on $k$-ary functions, i.e., $|D|=k$ in Remark \ref{rmk:S-on-D-ary} (do not confuse $|D|=k$ with the $k$ in $|D|^k$ there).

We use the following notation.  Let $(M, L, R)$ denote a partition of the input variables for  $(\TEP,h)$ where $M$ denotes the matrix at the root of the binary tree, $L$ and $R$ denote the left and right child, respectively. Note that both $L$ and $R$ correspond to inputs for $(\TEP,h-1)$. Recall $n_h$ denotes the input size for $(\TEP, h)$. For a subset $A \subseteq [n_h]$, we think of $A$ as a subset of input variables for $(\TEP,h)$ and write $A = (A_M, A_L, A_R)$, where $A_M = A \cap M$, $A_L = A \cap L$ and $A_R = A \cap R$. For notational simplicity, when the parameter $h$ is clear from the context, we use $\TEP_A$ to denote the matrix $(\TEP,h)_A$, and for $\alpha \in [k]^A$, we use $\TEP_\alpha$ to denote the subfunction $(\TEP,h)_{A,\alpha}$. We use $M_{ij} \in A_M$  to mean that $A_M$ contains the variable at entry $(i,j)$ of the root matrix $M$.

\begin{lemma}   \label{lem:equivalence}
Let 
	$h \ge 2$, 
	$A\subseteq [n_h]$. 
Suppose 
	$A=(\emptyset,A_L,A_R)$.  
Let 
	$\alpha= (\alpha_L,\alpha_R) \in [k]^{A} = [k]^{A_L} \times [k]^{A_R}$, 
	$\alpha' = (\alpha'_L,\alpha'_R) \in [k]^{A} = [k]^{A_L} \times [k]^{A_R}$. 
Then,
	\begin{align}  \label{eq:equivalence}
		&(\TEP,h)_{A,\alpha} = (\TEP,h)_{A,\alpha'}  \nonumber \\
		&\Longleftrightarrow
		\Big( (\TEP,h-1)_{A_L,\alpha_L}, (\TEP,h-1)_{A_R,\alpha_R} \Big)  = 
		\Big( (\TEP,h-1)_{A_L,\alpha'_L}, (\TEP,h-1)_{A_R,\alpha'_R} \Big).
	\end{align} 
\end{lemma}

\begin{proof}
The direction $\Longleftarrow$. Obvious.

The direction $\Longrightarrow$. 
	Assume for the sake of a contradiction the implication is not true. 
	Without loss of generality we may assume 
		$(\TEP,h-1)_{A_L,\alpha_L} \neq  (\TEP,h-1)_{A_L,\alpha'_L}$. 
	We will show
	\[
		(\TEP,h-1)_{A_L,\alpha_L} \neq  (\TEP,h-1)_{A_L,\alpha'_L}
		\Longrightarrow (\TEP,h)_{A,\alpha} \neq (\TEP,h)_{A,\alpha'}.
	\]
	Indeed, let 
		$\beta_L \in [k]^{L-A_L}$ 
		be such that 
		    $(\TEP,h-1)_{A_L,\alpha_L}(\beta_L) = i
		    \neq 
		    i' = (\TEP,h-1)_{A_L,\alpha'_L}(\gamma_L)$.
	Then, set $\beta_M \in [k]^{M}$ to be such that every entry in the $i$-th row equals to $i$, and every entry in the $i'$-th row equals to $i'$. 
	Choose $\beta_R  \in [k]^{R-A_R}$ arbitrarily. 
	Set
	    $\beta = (\beta_M,\beta_L,\beta_R)$.
	Then, 
	    $(\TEP,h)_{A,\alpha}(\beta) = i
	    \neq i'
	    = (\TEP,h)_{A,\alpha'}(\beta)$.
\end{proof}

\begin{lemma}   \label{lem:half-size}
For every $h\ge 2$, for every $\frac{n_h}{2} + 2^{h-2} \le \ell \le n_h$,  there exists a subset $A \subseteq [n_h]$ of size $|A| = \ell$ such that for every fixed $i\in [k]$, 
\[
\Pr_{\alpha} [\TEP_\alpha = i] \ge \frac{1}{2^{2^{h-1}-1} \cdot k},
\]
where $\alpha$ is chosen from $[k]^A$ uniformly at random. Here $\TEP_\alpha = i$ means the subfunction $(\TEP,h)_{A,\alpha}$ identically equals to $i$.
\end{lemma}

\begin{proof}
Suppose $k$ is even. 
Consider firstly the case $h=2$. Then $|A| \ge \frac{k^2 + 2}{2} + 1= \frac{k^2}{2}+2$. Choose $A$ such that $|A_M| = |A| - 2 \ge k^2/2$, $|A_R| = |A_L|  = 1$. In particular, both of the two leaves belong to $A$. Let $\alpha = (\alpha_M, \alpha_L, \alpha_R)\in [k]^{A_M} \times [k]^{A_L} \times [k]^{A_R}$. 
Let $Q = \{(r,s) \in [k] \times [k]: M_{rs} \in A_M\}$. Let $\alpha$ be chosen from $[k]^A$ uniformly at random. Then, for every fixed $i \in [k]$, 
\begin{equation}   \label{eq:partition}
\Pr_{\alpha} [\TEP_\alpha = i]
= \sum_{(r,s)\in Q} \Pr_{\alpha_L}[\alpha_L = r] \times \Pr_{\alpha_R}[\alpha_R = s] \times \Pr_{\alpha_M}[M_{rs} = i]
= \frac{|Q|}{k^3}
= \frac{|A_M|}{k^3}
\ge \frac{1}{2k}.
\end{equation}

We use induction to prove the general case. Specifically,  we choose $A$ such that, 
\begin{enumerate}[(1)]
\item $|A_M| \ge k^2/2$;
\item choose $A_L$ so that all the leave variables in $L$ are chosen, and for every matrix node in $L$, $A_L$ contains at least a half of its entries;
\item choose $A_R$ similarly as $A_L$. 
\end{enumerate}
The above choice for $A$ is feasible because $|A| \ge \frac{n_h}{2} + 2^{h-2}$. Observe that the choice of $A_L$  and $A_R$ allows induction on height $h-1$. Hence, by the same calculation as \eqref{eq:partition} and by induction on height $h-1$, one has that for every fixed $i\in [k]$, 
\[
\Pr_{\alpha} [\TEP_\alpha = i] 
\ge |A_M| \times \frac{1}{2^{2^{h-2}-1} \cdot k} \times \frac{1}{2^{2^{h-2}-1} \cdot k} \times \frac{1}{k}
\ge \frac{1}{2^{2^{h-1}-1} \cdot k}. 
\] 
Finally, it is easy to see that the same argument also applies  when $k$ is odd. We omit the details.  
\end{proof}

\begin{theorem}   \label{thm:hat-s-TEP}
For $k\ge 2$ and $1 \le h\le \log k$, one has 
\[
k \le \widehat{S}(\TEP,h) \le \frac{2^{2^{h-1}+1}}{3} \cdot k.
\]
In particular, 
    $\widehat{S}(\TEP,\log \log k) \le k^2/3$.
\end{theorem}

\begin{proof}
By definition, $\widehat{S}(\TEP,h) = \max_{1 \le \ell \le n_h} \min_{|A| = \ell} \frac{k^{|A|}}{\mult((\TEP,h)_A)}$.

The lower bound. Choosing $\ell = n_h$ and $A = [n_h]$ implies that $\mult(\TEP_A) = k^{n_h - 1}$, hence $\widehat{S}(\TEP,h) \ge k$. 

The upper bound. We will use induction on $h$. The base case when $h=1$ is clear: $\widehat{S}(\TEP,1) = k \le 4k/3$. Assume now $h\ge 2$ and the claim is true for $h-1$.  By definition of $\widehat{S}$, we will show that for each $1 \le \ell \le n_h$, there exists a subset $A \subseteq [n_h]$ such that $|A| = \ell$ and $\mult(\TEP_A) \ge \phi(h) k^{|A|-1} = \phi(h) k^{\ell-1}$, where $\phi(h) = 3/(2^{2^{h-1}+1})$. We consider three cases.

\begin{itemize}
\item Case 1: $\ell \le n_{h-1}$. 
	Choose $A$ such that 
		$A_M = \emptyset$, 
		$|A_L| = \ell$ and 
		$A_R  = \emptyset$,
	i.e., 
		$A = A_L$.
	
	Apply Lemma \ref{lem:equivalence} in the case when $A_R  = \emptyset$, 
	one has that for every $\alpha, \alpha' \in [k]^{A} = [k]^{A_L}$,
	\[
		(\TEP,h)_{A,\alpha} = (\TEP,h)_{A,\alpha'}
		\Longleftrightarrow
		(\TEP,h-1)_{A_L,\alpha} = (\TEP,h-1)_{A_L,\alpha'}	
	\]
	Hence,
		$m\Big( (\TEP,h)_{A} \Big) = m\Big( (\TEP,h-1)_{A_L} \Big)$.
	By induction, there exists $A_L$ such that 
		$m\Big( (\TEP,h-1)_{A_L} \Big) \ge \phi(h-1) k^{\ell - 1}$. 
	Hence, 
		$m\Big( (\TEP,h)_{A} \Big) \ge  \phi(h-1) k^{\ell - 1} \ge \phi(h) k^{\ell - 1}$.

\item Case 2: $n_{h-1} < \ell \le n_{h-1}+ k(k-1)$. Choose $A$ to be of size $\ell$ such that,
\begin{enumerate}[(1)]
\item $A_M$ is a subset of the entries in the first $k-1$ rows of $M$;
\item $|A_L|=n_{h-1}$;
\item $A_R = \emptyset$. 
\end{enumerate}
Clearly, the above choice is feasible. Let 
\[
\cF = \{\alpha = (\alpha_M, \alpha_L, \alpha_R)\in [k]^{A_M} \times [k]^{A_L} \times [k]^{A_R}: (\TEP,h-1)_{A_L,\alpha_L} = k\}. 
\]
Since $|A_L|=n_{h-1}$, we have $\Pr_{\alpha_L}[(\TEP,h-1)_{A_L,\alpha_L} = k] = 1/k$.  Hence, $|\cF| = k^{|A| - 1}$. The choice of $A$ implies that $(\TEP,h)_{A,\alpha} = (\TEP,h)_{A,\beta}$ whenever $\alpha, \beta \in \cF$.  Hence, $m\Big( (\TEP,h)_A \Big) \ge |\cF| = k^{|A| - 1} > \phi(h) k^{|A| - 1}$.

\item Case 3: $\ell > n_{h-1} + k(k-1)$. Choose $A$ to be of size $\ell$ such that,
\begin{enumerate}[(1)]
\item $k(k-1) - 2^{h-2} < |A_M|  \le k^2$; 
\item $|A_L| \ge \frac{n_{h-1}}{2} + 2^{h-3}$;
\item $|A_R| \ge \frac{n_{h-1}}{2} + 2^{h-3}$. 
\end{enumerate}
Clearly, the above choice is feasible. Moreover, $|A_M| \ge k^2 - k - 2^{h-2} \ge k^2 - 5k/4$ as $h \le \log k$. Specification on  $A_L$ an $A_R$ will be given by Lemma \ref{lem:half-size}. Indeed, by Lemma \ref{lem:half-size}, there exist choices of $A_L$ and $A_R$ such that for every $(i,j) \in [k] \times [k]$, 
\begin{equation}      \label{eq:bounds}
\Pr_{\alpha_L} [(\TEP,h-1)_{A_L,\alpha_L} = i] \ge \frac{1}{2^{2^{h-2}-1} \cdot k},
\quad
\Pr_{\alpha_R} [(\TEP,h-1)_{A_R,\alpha_R} = j] \ge \frac{1}{2^{2^{h-2}-1} \cdot k}.
\end{equation} 
Let $Q=\{(i,j) \in [k] \times [k]: M_{ij} \in A_M\}$. Similar to \eqref{eq:partition}, apply \eqref{eq:bounds}, 
\begin{align*}
&\Pr_\alpha[(\TEP,h)_{A,\alpha} = 1] \\
&=
\sum_{(i,j) \in Q} \Pr_{\alpha_L} [(\TEP,h-1)_{A_L,\alpha_L} = i] \times \Pr_{\alpha_R} [(\TEP,h-1)_{A_R,\alpha_R} = j] \times \Pr_{\alpha_M}[M_{ij} = 1]  \\
&\ge \frac{|A_M|}{2^{2^{h-1}-2} \cdot k^3}
\ge \frac{1- \frac{5}{4k}}{2^{2^{h-1}-2}} \cdot \frac{1}{k}
\ge \frac{3}{2^{2^{h-1}+1}} \cdot \frac{1}{k}
\end{align*}
Hence, $\mult((\TEP,h)_A) \ge \frac{3}{2^{2^{h-1}+1}} \cdot  k^{|A|-1} = \phi(h) \cdot  k^{|A|-1}$.
\end{itemize}
In all three cases we have verified the existence of the desired $A$, the proof is completed.
\end{proof}

\subsection{$S$ on $\TEP$}    \label{sec:S-TEP}

Let 
    $S((\TEP,h), \ell) = \min_{A \subseteq [n_h],|A| = \ell} \nrows((\TEP,h)_A)$.
Then,
    $S(\TEP,h) = \max_{1 \le \ell \le n_h} S((\TEP,h), \ell)$.

\begin{theorem}  \label{thm:S-TEP}
    \begin{enumerate}[(1)]
        \item $S(\TEP,2) = k^2$
            and is  achieved at either $\ell = k+1$ or $\ell = k+2$, i.e.,
                $S((\TEP,2),k+1) = S((\TEP,2),k+2) = k^2$;
                
        \item $S(\TEP,3) \ge k^3/16$ for $k\ge 4$;
        
        \item $S(\TEP,h) \le k^h - k^{h-2}(k-2) < k^h$ for $h \ge 3$.
    \end{enumerate}
\end{theorem}  

Theorem \ref{thm:S-TEP}-(1) shows that the value of $S(f)$ is not necessarily achieved at $|A| = n/2$. The complete  proof is lengthy, we provide it in  Appendix \ref{sec:appendix}.

\section{The read-once non-deterministic BP lower bound of satisfiable Tseitin formulas via $\widehat{C}$ : a short proof} \label{sec:Tseitin}

In a line of recent works \cite{TseitinOBDD2017,Tseitin2017,Tseitin2019},  BP ($\OBDD$ and $\nBP_1$) sizes of satisfiable Tseitin formulas are studied for applications in proof complexity. 
Let $\kappa(H)$ denote the number of connected components of graph $H$.
Define 
    $\kappa_G(\ell) = \max_{H \le G, |E(H)| = \ell} \kappa(H)$,
where  $H\le G$ means $H$ is a subgraph of $G$. 
\cite{Tseitin2017,Tseitin2019} proved that if $\kappa(G) = 1$, then
    $\nBP_1(\TS_{G,c}) \ge \max_{1\le \ell \le m} 2^{n - \kappa_G(\ell) - \kappa_G(m-\ell) + 1}$,
and used this as a starting point to prove $\nBP_1$ lower bounds for satisfiable Tseitin formulas. 
Below we give a short proof  using $\widehat{C}$ for arbitrary $\kappa(G)$.   We need  the following fact. 

\medskip

{\noindent \bf Fact (\cite[Lemma 2]{Tseitin2017}).} $|\TS_{G,c}^{-1}(1)| = 2^{m-n+\kappa(G)}$.

\begin{theorem}     \label{thm:Tseitin}
Let $\TS_{G,c}$ be a satisfiable Tseitin formula. Then,
    \[
        \widehat{C}(\TS_{G,c}) \ge \max_{1 \le \ell \le m} 2^{n - \kappa_G(\ell) - \kappa_G(m-\ell) + \kappa(G)}.
    \]
\end{theorem}

\begin{proof}
Fix an
    $1 \le \ell \le m$, 
consider an arbitrary $A \subseteq [m]$ of size $|A| = \ell$, let $B = [m] - A$. 
We view $A \subseteq E$, so $A \cup B = E$. 
Consider an $A$-rectangle 
    $(p,q): \{0,1\}^A \times \{0,1\}^{B} \to \{0,1\}$,
that is,
    $(p,q)(a,b) = p(a) \land q(b)$
for every $(a,b) \in \{0,1\}^A \times \{0,1\}^{B} = \{0,1\}^E$. 
Suppose further the rectangle satisfies 
    $(p,q)(a,b) = 1$ 
implies $\TS_{G,c}(a,b) = 1$ 
for every     $(a,b)$.
Let 
    $P = p^{-1}(1) \subseteq \{0,1\}^A$
and
    $Q = q^{-1}(1) \subseteq \{0,1\}^{B}$. 
Then, it suffices to show $|P| \times |Q|$  is ``small''.

Let $v \in V$. Let $E(v)$ denote the set of edges in $E$ that are incident to $v$, let 
    $E(A,v) = E(v) \cap A$ and 
    $E(B,v) = E(v) \cap B$.
Then,
    $E(v) = E(A,v) \cup E(B,v)$.
Given $b \in \{0,1\}^{B}$, by an abuse of notation,  we define a labelling function  
     \[
        b: V \to \{0,1\}, \quad
        v \mapsto 
        \begin{cases}
            0, &\quad E(B,v) = \emptyset, \\
            \sum_{e \in E(B,v)} b_e \mod 2, &\quad \text{otherwise}.
        \end{cases}
    \]
Then, for 
    $x=(a,b) \in \{0,1\}^A \times \{0,1\}^{B} = \{0,1\}^E$,
we have
    \begin{equation}  \label{eq:decomp}
        \sum_{e \in E(v)} x_e
        = \sum_{e \in E(A,v)} a_e + \sum_{e \in E(B,v)} b_e
        \equiv b(v) + \sum_{e \in E(A,v)} a_e \mod 2.
    \end{equation}
Now, fix an arbitrary $b^* \in Q$.  By \eqref{eq:decomp}, for every $a\in P$, we have 
    \begin{align*}
        (p,q)(a,b^*) = 1  
        \Longrightarrow \TS_{G,c}(a,b^*) = 1
        &\Longrightarrow
        \forall\ v \in V,\  
        b^*(v) + \sum_{e \in E(A,v)} a_e \equiv c(v) \mod 2 \\
        &\Longrightarrow
        \forall\ v \in V,\  
        \sum_{e \in E(A,v)} a_e \equiv c(v) - b^*(v) \mod 2.
    \end{align*}
Let $H_A = (V, A)$ be the subgraph of $G$ given by edges in $A$. The above and the {\bf Fact} imply that 
    \[
        \forall\ a \in P,\ \TS_{H_A,c - b^*}(a) = 1
        \Longrightarrow
        |P| \le |\TS_{H_A,c - b^*}^{-1}(1)| = 2^{|A| - n + \kappa(H_A)}
        \le 2^{\ell - n + \kappa_G(\ell)}.
    \]
By symmetry, 
    $|Q| \le 2^{m-\ell - n + \kappa_G(m-\ell)}$.
Hence,
    $|P| \times |Q| \le 2^{m - 2n + \kappa_G(\ell) + \kappa_G(m-\ell)}$.
The claim follows by applying the {\bf Fact} again to $\TS_{G,c}$.
\end{proof}

\section{$\widehat{S}$ and $\widehat{C}$ on G{\'a}l-type functions, and related blocking sets}   \label{sec:Gal}

G{\'a}l's original function \cite{gal1997simple} was defined using projective planes, while a closely related function the Bollig-Wegener function \cite{bollig1998very} was defined from representing numbers in a prime basis. In Section \ref{sec:funcs} we have defined both functions with respect to a given bipartite graph. This definition makes the connection between the two readily apparent. Indeed, by definition in Section \ref{sec:funcs}, for every $x \subseteq A$,
\begin{equation}    \label{eq:Gal-to-BW}
    \GAL_G(x) = 1 
    \Longleftrightarrow
    \BW_G(x,y) = 1, \forall\ \emptyset \neq y \subseteq B
    \Longleftrightarrow
    \BW_G(x,\{b\}) = 1, \forall\ b \in B.
\end{equation} 
Furthermore, our definition also naturally allows a possible further approach to tackle the $\BPone$ vs weight problem, as we will see shortly.
Below we discuss $\widehat{S}$ and $\widehat{C}$ on generalized G{\'a}l's function and Bollig-Wegener function, and related mathematical problems.

\subsection{The point-line incidence graph over finite fields}

Here we specify a bipartite graph to instantiate the GAL and BW functions we will study.
Let $\F_q$ be a finite field of order $q$, we use the notation $\F_p$ to denote the case when $p$ is a prime and hence $\F_p$ is a prime field. We associate every $(i,j) \in \F_q^2$ a unique \emph{non-vertical} line $\ell_{(i,j)} \subseteq \F_q^2$ given by 
\begin{equation}   \label{eq:non-v-line-by-ij}
    \ell_{(i,j)} = \{(t, i+ jt): t \in \F_q\}.
\end{equation}
Note that the ``direction'' of the line $\ell_{(i,j)}$ is $(1,j)$, hence is ``non-vertical'', whereas we think of the direction $(0,1)$ as the \emph{vertical} direction. Let $A=B=\F_q^2$. Consider the bipartite graph $G_q(A\cup B, E)$ defined by the point-line incidence relation. Specifically, there is an edge between $a\in A$ and $b \in B$ if and only if $a \in \ell_b$. In other words, we think of $A$ as the set of $q^2$ points of $\F_q^2$, and $B$ as the set of $q^2$ non-vertical lines in $\F_q^2$. It is easy to verify the following.

\begin{lemma}   \label{lem:G_q}
    The bipartite graph $G_q$ is $q$-regular and $K_{2,2}$-free.
\end{lemma}

\begin{definition}  \label{def:blocking-set}
    A subset of points $S \subseteq \F_q^2$  is called a \emph{blocking set} if and only if it intersects every non-vertical line, 
    it is called a \emph{minimal blocking set} if no subset of $S$ is a blocking set.
\end{definition}

\begin{remark}
    The usual definition of blocking sets in an affine plane requires to intersect \emph{all} lines, not just the non-vertical lines, see e.g., \cite{brouwer1978blocking}. Blocking sets under this definition in both affine and projective spaces have been widely studied in the past several decades, see \cite{hirschfeld1998projective,blokhuis2002combinatorial}, most studies focus on classifying minimal blocking sets. Our definition is slightly different, we will also ask a different question (Question 1) in Section \ref{sec:Gal-BW-SChat}.
\end{remark}

Using the bipartite graph $G_q(A\cup B, E)$,  a subset $S \subseteq A$ is a blocking set if and only if $\Nb(S) = B$. It is also not hard to show that $S \subseteq \F_p^2$ is a blocking set if and only if the polynomial $P_S(x,y) \in \F_p[x,y]$
    \begin{equation}    \label{eq:poly-blockingset}
        P_S(x,y) = \prod_{(a,b) \in S} (x+ay - b)
    \end{equation}
is identically $0$ on $\F_p^2$.

We will consider $\GAL_{G_q}$ and $\BW_{G_q}$. In this case, $\GAL_{G_q}(x) = 1$ if and only if $x$ is a blocking set in $\F_q^2$. G{\'a}l in \cite{gal1997simple} defined her function similar to the one given above, but in the projective plane  $\PG(2,q)$ (see the detail in \cite{gal1997simple}), instead of the affine plane $\F_q^2$. Bollig and Wegener defined their function in \cite{bollig1998very}  in a different language, and remarked that their function $\BW$ ``is somehow similar to the construction in \cite{gal1997simple}''.  By  phrasing both functions over $G_q$, this connection is shown in \eqref{eq:Gal-to-BW}.

\subsection{$\widehat{S}$ and $\widehat{C}$}    \label{sec:Gal-BW-SChat}

G{\'a}l in \cite{gal1997simple} showed that her function defined using the projective plane $\PG(2,q)$ is $q$-mixed, hence has exponential $\BPone$ complexity. Below we generalize G{\'a}l's result to arbitrary bipartite graphs satisfying appropriate conditions, by a similar argument as  \cite{gal1997simple}. 

\begin{figure}[h!]
\centering
        \includegraphics[scale=.9]{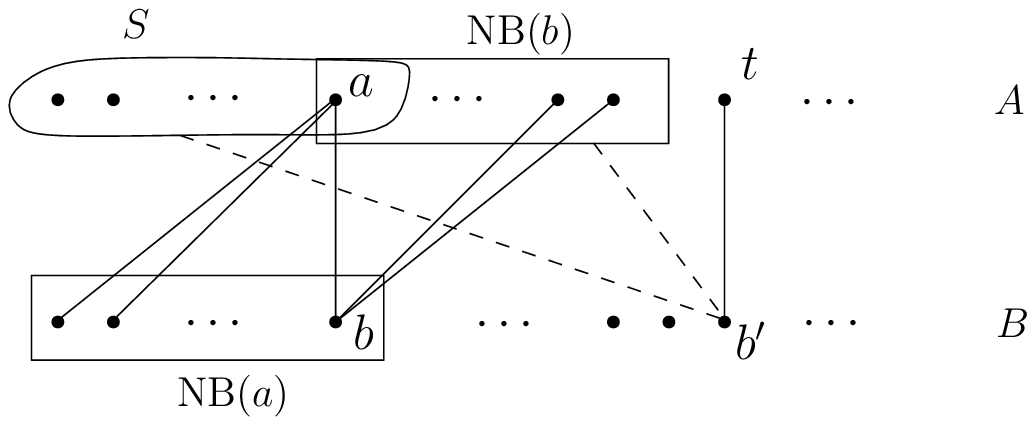}
        \caption{The illustration for Theorem \ref{thm:Gal-mixed}.}
        \label{Fig:alternative_pf_GAL}
\end{figure}

\begin{theorem} \label{thm:Gal-mixed}
    Let $G=(A\cup B, E)$ be a bipartite graph that is $r$-regular and $K_{2,s}$-free for some $s \ge 2$.  Then, 
        $\widehat{S}(\GAL_G) \ge 2^{\floor{(r-1)/(s-1)}}$.
\end{theorem}

\begin{proof}
    By Lemma \ref{lem:mixed-to-S-hat}, it suffices to show that $\GAL_G$ is $\floor{(r-1)/(s-1)}$-mixed. Let $S \subseteq A$ be an arbitrary subset of size $\floor{(r-1)/(s-1)}$, let $\alpha \neq \beta$ be two subsets of $S$ and assume $a \in \alpha \backslash \beta$, see Figure \ref{Fig:alternative_pf_GAL}. We need to show that there exists a subset $\gamma \subseteq A \backslash S$ such that $\Nb(\alpha \cup \gamma) = B$
    but 
    $\Nb(\beta \cup \gamma) \neq B$.  
    
    Since $G$ is $K_{2,s}$-free and $r$-regular, and $|S|=\floor{(r-1)/(s-1)}$, 
    \[
        |\Nb(\beta) \cap \Nb(a)| \le (s-1) |\beta|
        < (s-1) |S| \le r-1 < |\Nb(a)|
        \Longrightarrow
        \exists\ b \in \Nb(a) \backslash \Nb(\beta).
    \]
    Take $\gamma = A \backslash (S \cup \Nb(b))$. Then, 
        $b \not\in (\Nb(\beta) \cup \Nb(\gamma))$,
        i.e.,
        $\Nb(\beta \cup \gamma) \neq B$.
    To show 
        $\Nb(\alpha \cup \gamma) = B$, 
    consider an arbitrary $b' \not\in \Nb(a)$. Again, since $G$ is $K_{2,s}$-free and $r$-regular and $|S|=\floor{(r-1)/(s-1)}$,
    \begin{align*}
        |(S \cup \Nb(b)) \cap \Nb(b')|
        &\le |(S \backslash \{a\}) \cap \Nb(b') | + |\Nb(b) \cap \Nb(b') |      \\
        &\le (s-1)|S \backslash \{a\}| + (s-1) 
        \le r-1
        < |\Nb(b')|.
    \end{align*}
    Hence, there exists 
        $t \in \gamma \cap \Nb(b')$.
    This together with 
        $\Nb(a) \subseteq \Nb(\alpha)$
    shows
        $\Nb(\alpha\cup \gamma) = B$.
\end{proof}

\begin{corollary}
    $\BPone(\GAL_{G_q}) \ge 2^{q-1}$.
\end{corollary}

\begin{proof}
    Apply Proposition \ref{prop:BP1-LB}, Lemma \ref{lem:G_q} and Theorem \ref{thm:Gal-mixed}.
\end{proof}

Similar to G{\'a}l's $\BPone$ lower bound \cite{gal1997simple}, Bollig and Wegener in \cite{bollig1998very} also showed that $\BPone(\BW_{G_q})$ has an exponential lower bound. In \cite{bollig1998very} they pointed out that $\BW_{G_q}$ is not mixed, here we strengthen this by showing that in fact $\widehat{S}(\BW_{G_q})$ is small.

\begin{figure}[h!]
\centering
        \includegraphics[scale=.9]{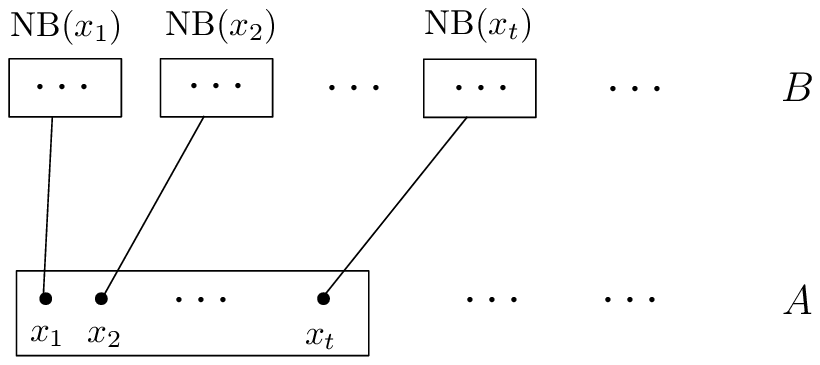}
        \caption{The illustration for Theorem \ref{thm:BW-S-hat-small}.}
        \label{Fig:BW-S-hat-small}
\end{figure}

\begin{theorem} \label{thm:BW-S-hat-small}
    $\widehat{S}(\BW_{G_q}) = O(1)$.
\end{theorem}

\begin{proof}
    Let $k_t=t(q+1)$. We first show for $1 \le t \le q$, there exist $S\subseteq A \cup B$ with size $|S|=k_t$ such that $\mult((\BW_{G_q})_S)$ is large. 
    Let $V = \{x_1, \ldots, x_q\} \subseteq A$ be a subset of $q$ points that form a vertical line. Then, $\Nb(x_i) \cap \Nb(x_j) = \emptyset$ and $V$ is a minimal blocking set. 
    Let 
        $T = \{x_1, \ldots, x_t\}  \subseteq V$.
    Choose
        $S = T \cup \Nb(T)$,
    so 
        $|S| = t(q+1) = k_t$,
    see Figure \ref{thm:BW-S-hat-small}.
    Let $H(S, E(S)) \le G_q(A\cup B, E)$ denote the induced subgraph by $S$.
    Let 
        $(\alpha,\beta) \in \{0,1\}^T \times \{0,1\}^{\Nb(T)}$.
    With this notation, one has 
    \begin{equation}    \label{eq:reduce-to-H}
        (\BW_{G_q})_{S,(\alpha,\beta)} = 1
            \Longleftrightarrow
            \BW_{H}(\alpha,\beta) = 1.
    \end{equation}
    Let $P_t$ denote the probability that $\BW_{H}(\alpha,\beta) = 0$ where $(\alpha,\beta)$ is chosen uniformly at random, i.e., the probability that there are no edges between $\alpha$ and $\beta$ in $H$.
    \[
        P_t = \sum_{i=0}^t {t \choose i} 2^{-t} 2^{-qi} 
        = (2^{-1} + 2^{-(q+1)})^t 
        \le 5/8
        \Longrightarrow
        \Pr[\BW_{H}(\alpha,\beta) = 1] \ge 3/8.
    \]
    By \eqref{eq:reduce-to-H}, this implies 
        $\mult((\BW_{G_q})_S) \ge 2^{k_t} \cdot 3/8 $
    as desired.
    With slight modification, the other values of $1 \le k \le 2q^2$ can all be handled, we omit the details. 
\end{proof}

Next we discuss $\widehat{C}(\GAL_{G_q})$, it is best to look at this problem within a background. 
Using the terminology from \cite{jukna1999p}, define the \emph{weight} of $f: \{0,1\}^n \to \{0,1\}$ to be
    \begin{equation}    \label{eq:def-weigth}
        w(f) = \DNFS(f) + \CNFS(f),
    \end{equation}
where $\DNFS(f)$ and $\CNFS(f)$ denote the DNF and CNF sizes of $f$, respectively.
    
\smallskip
{\noindent \bf Open Problem (\cite{bollig1998very,jukna1999p,wegener2000branching}).} Does there exist a Boolean function  $f$ such that 
    $\BPone(f)$
is superpolynomially larger than
    $w(f)$? 
\smallskip

This problem has been open for more than two decades. If we replace $\BPone$ by  $\OBDD$, we could deduce separations in both sides. Let $\DTS(f)$ denote the decision tree size of $f$. 

\begin{proposition} \label{prop:obdd-Q4}
    $w(\NAND_n) \ge \exp(\Omega(\sqrt{n/\log n}))$ but $\OBDD(\NAND_n) = O(n)$.
        
    $w(\ISA_n) \le \DTS(\ISA_n) = O(n^2)$ but $\OBDD(\ISA_n) = \exp(\Omega(n/\log n))$.
\end{proposition}

\begin{proof}
In \cite{ehrenfeucht1989learning} it shows (see also a proof from  \cite{jukna2012boolean})
    $\DTS(f) \le \exp(O((\log n) \cdot \log^2 w(f)))$.
By \cite{jukna1999p},
    $\DTS(\NAND_n) \ge 2^{\Omega(n)}$.
These two together implies the lower bound for $w(\NAND_n)$.  Theorem 4.3.3 from \cite{wegener2000branching} shows that $\OBDD(\ISA_n) \ge \exp(\Omega(n/\log n))$. The upper bounds are easy to see.
\end{proof}

It is easy to see that 
    $\CNFS(\GAL_{G_q}) = O(q^3)$
and 
    $\DNFS(\BW_{G_q}) = O(q^3)$.
Hence, it is interesting to ask what is $\DNFS(\GAL_{G_q})$? It is not hard to see that
$\DNFS(f) \ge \widehat{C}(f)$. Hence, for the purpose of a lower bound, it suffices to show $\widehat{C}(\GAL_{G_q}, n/2)$ is large, where $n=q^2$. This raises naturally the following question.

\smallskip
{\bf \noindent Question 1.} Let $M \subseteq \F_q^2$ be an arbitrary subset of size $q^2/2$. Is the number of blocking sets inside $M$ always exponentially (or superpolynomially) smaller than $2^{q^2/2}$?
\smallskip

Indeed, it is not hard to see that if there is an $M \subseteq \F_q^2$ of size $n/2=q^2/2$ with $t$ blocking sets inside, then $\GAL_{G_q}$ contains  a $1$-monochromatic $n/2$-rectangle of size $t 2^{p^2/2}$. If one wishes to show $\widehat{C}(\GAL_{G_q},n/2)$ is large by showing  that every $1$-monochromatic $n/2$-rectangle is small, then answering Question 1 is crucial. The following simple fact is a  contrast to Question 1.

\begin{proposition}     \label{prop:all-are-blocking-sets}
    As $q\to \infty$, almost all subsets in $\F_q^2$ are blocking sets.    
\end{proposition}

\begin{proof}
    By definition, every non-blocking-set is a subset of $\F_q^2 \backslash \ell$ for some (non-vertical) line $\ell$. As there are $q^2$ such lines,
        the number of non-blocking-set
        $\le q^2 \cdot 2^{q^2 - q}$.
    This implies the number of blocking sets
        $\ge 2^{q^2} (1 - q^2/2^q) \to 2^{q^2}$,
    as  $q \to \infty$.
\end{proof}

One potential way towards answering Question 1 might be the following.

\smallskip
{\bf \noindent Question 2.} Classify all minimal blocking sets in $\F_q^2$.
\smallskip

\subsection{Some results on minimal blocking sets in $\F_p^2$}  \label{sec:minimal-BS}

Here we give some results for Question 2. Structures of minimal blocking sets (MBS for short) might also be of independent interest. 
For simplicity throughout this section we work with $\F_p^2$ where $p$ is a prime. When we say lines we mean non-vertical lines unless specified otherwise. We say two lines are parallel if they are two distinct lines with the same direction.

\begin{figure}[h!]
\centering
        \includegraphics[scale=.9]{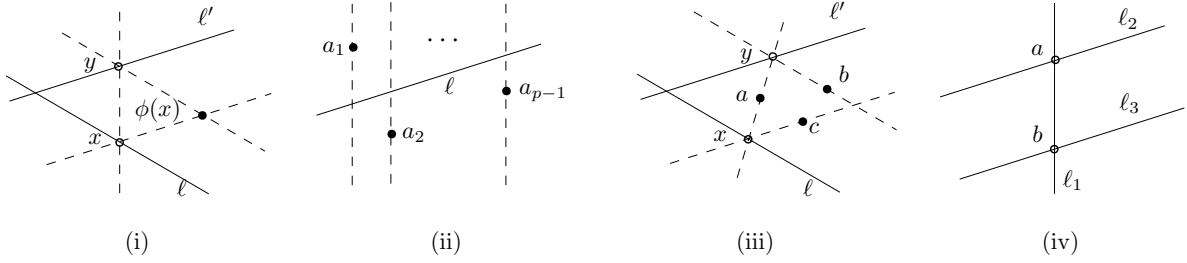}
        \caption{The illustration for Theorem \ref{thm:MBS}.}
        \label{Fig:MBS-types}
\end{figure}

\begin{theorem} \label{thm:MBS}
    MBS in $\F_p^2$ have size at least $p$. Furthermore, 
    \begin{enumerate}[(1)]
        \item There are exactly $p$ MBS of size $p$, each of which is a \emph{vertical} line.
        
        \item There are no  MBS of size $p+1$.
        
        \item Let $\ell$ and $\ell'$ be two intersecting lines. Let $x\in \ell \backslash \ell'$. Let $y \in \ell'$ be the unique point in $y$ such that the line determined by $x$ and $y$ is vertical. Let $\ell_y$ be the  unique line parallel to $\ell$ and passing through $y$. Let $\ell'_x$ be the  unique line parallel to $\ell'$ and passing through $x$.
        Define
            $\phi(x) = \ell_y \cap \ell'_x$.
        Then, 
            $\ell \cup \ell' \cup \{\phi(x)\} \backslash \{x,y\}$ 
        is an MBS of size $2p-2$, see Figure \ref{Fig:MBS-types}-(i).
        
        \item Every pair of two intersecting lines is an MBS of size $2p-1$.
        
        \item Let $\ell$ be a line. Let 
            $a_1, \ldots, a_{p-1}$ 
        be $p-1$ points not in $\ell$, such that no two of them are in the same vertical line, and such that every line parallel to $\ell$ passes through some $a_i$. Then, 
            $\ell \cup \{a_1, \ldots, a_{p-1}\}$ 
        is an MBS of size $2p-1$,  see Figure \ref{Fig:MBS-types}-(ii).
        
        \item Let $\ell$ and $\ell'$ be two intersecting lines, let $x\in \ell$, let $y \in \ell'$ be such that the line determined by $x$ and $y$ is \emph{not} vertical. Let $\ell_y$ and $\ell'_x$ be the two lines that are parallel to $\ell$ and $\ell'$, respectively. Let 
            $a \neq x, y$ 
        be a point in the line determined by $x$ and $y$. 
        Let 
            $b \in \ell_y$, $b\neq y$ and $b \neq \ell_y \cap \ell'_x$. 
        Let 
            $c \in \ell'_x$, $c\neq x$ and $c  \neq \ell_y \cap \ell'_x$.  
        Then,
            $\ell \cup \ell' \cup \{a,b,c\} \backslash \{x,y\}$ 
        is an MBS of size $2p$, see Figure \ref{Fig:MBS-types}-(iii).
        
        \item Let $\ell_1$ be a vertical line. Let $\ell_2$ and $\ell_3$ be two parallel lines intersecting with $\ell_1$ at $a$ and $b$, respectively. 
        Then,
            $\ell_1 \cup \ell_2 \cup \ell_3 \backslash \{a,b\}$
        is an MBS of size $3p-4$, see Figure \ref{Fig:MBS-types}-(iv).
    \end{enumerate}
\end{theorem}

Most of the statements in Theorem \ref{thm:MBS} are not hard to prove. For example, the MBS has size at least $p$ can be deduced via a degree argument by the polynomial characterization \eqref{eq:poly-blockingset} (of course, it can also be proved directly). We find the proof for (5) is particularly interesting, and the lemmas used to prove it might be of independent interest. So we give its proof below.

\subsubsection{A lemma of intersecting points in $\F_p^2$}  \label{sec:intersecting-points}

\begin{lemma}   \label{lem:identity}
    Let $p$ be a prime, let $x_1, \ldots, x_{p-1} \in \{1, \ldots, p-1\}$ be distinct. Then, 
        $\prod_{1 \le i < j \le p-1} (i x_j - j x_i)  \equiv 0 \mod p$.
\end{lemma}

\begin{proof}
    For $1 \le i \le p-1$, let 
        $y_i = x_i / i \in \{1,\ldots, p-1\}$.
    Writing 
        $i x_j - j x_i= i j ( y_j - y_i)$, 
    one observes that in $\F_p$,
        $\prod_{1 \le i < j \le p-1} (i x_j - j x_i) = 0$
    is equivalent to
        $\prod_{1 \le i < j \le p-1} (y_j - y_i) = 0$.
    So it suffices to show that $y_i =y_j$ for some $i \neq j$. Assume otherwise, i.e.,  $y_1, \ldots, y_{p-1} \in \{1,\ldots, p-1\}$ are all distinct. Then, on one hand, by Wilson's theorem,
        $\prod_{i=1}^{p-1} y_i = (p-1)! \equiv  -1 \mod p$.
    On the other hand, since $x_i$ are all distinct,
        $\prod_{i=1}^{p-1} y_i
        = \prod_{i=1}^{p-1} \frac{x_i}{i}
        = \frac{\prod_{i=1}^{p-1} x_i}{ \prod_{i=1}^{p-1} i}
        = \frac{(p-1)!}{(p-1)!} = 1 \mod p$,
    contradicting to the previous equation. 
\end{proof}

Below in Lemma \ref{lem:intersecting-points-property} lines mean arbitrary lines, i.e., either vertical or non-vertical. 

\begin{lemma}   \label{lem:intersecting-points-property}
    Let 
        $\ell_1, \ldots, \ell_p$ 
    be $p$ parallel lines in $\F_p^2$, 
    let 
        $\ell'_1, \ldots, \ell'_p$ 
    be $p$ parallels in a different direction. 
    For $1\le i \le p$, let 
        $a_i = \ell_i \cap \ell'_i$. 
    Then, for every point $a_i$, there exists $a_j, a_k$, $j,k\neq i$ and $j\neq k$, such that these three points $a_i, a_j, a_k$ are colinear.
\end{lemma}

\begin{proof}
    Note that there are $p+1$ different directions, i.e., 
        $(0,1)$ and $(1,d)$ for $d \in \F_p$.
        
    Consider the case where the two directions are     $(1,d)$ and $(1,d')$, where $d \neq d'$,
    for lines 
        $\ell_1, \ldots, \ell_p$ and
        $\ell'_1, \ldots, \ell'_p$, respectively.
    By \eqref{eq:non-v-line-by-ij}, the lines $\ell_i$ can be parametrized as 
    \begin{equation}   \label{eq:ell_i}
        \ell_i = \{(t, m_i + d t): t \in \F_p\}, \quad 
        i= 1, \ldots, p,
        \quad
        \text{ where }
        m_1, \ldots, m_p \in \F_p
        \text{ are all distinct}.
    \end{equation}
    Similarly, lines $\ell'_i$ can be parametrized as
    \begin{equation}   \label{eq:ell-prime_i}
        \ell'_i = \{(t, n_i + d' t): t \in \F_p\}, \quad 
        i= 1, \ldots, p,
        \quad
        \text{ where }
        n_1, \ldots, n_p \in \F_p
        \text{ are all distinct}.
    \end{equation}
    By \eqref{eq:ell_i} and \eqref{eq:ell-prime_i}, we have 
    \begin{equation}    \label{eq:a_i}
        a_i = (t_i, m_i + d t_i), \quad
        \text{ where }
        t_i = \frac{m_i - n_i}{d' - d},
        \quad 
        i= 1, \ldots, p.
    \end{equation}
    Note that since 
        $d' \neq d$,
    $t_i \in \F_p$ is well-defined.
    
    Without loss of generality, we may assume the point that is fixed is $a_1$, and we wish to show there exist $2 \le j < k \le p$ such that 
        $a_1, a_j, a_k$
    are in the same line. 
    This is equivalent to showing the 
        $p-1$
    directions
        $a_2 - a_1, \ldots, a_p - a_1$ 
    are not all distinct. 
    By \eqref{eq:a_i}, 
        $a_i - a_1 = (t_i - t_1, m_i - m_1 + d(t_i - t_1))$.
    The directions 
        $a_j - a_1$ and $a_k - a_1$
    are the same if and only if 
    \[
        \det(A_{jk}) = 0, 
        \quad
        \text{ where matrix }
        A_{jk} =
        \begin{pmatrix}
            t_j - t_1 & m_j - m_1 + d(t_j - t_1) \\
            t_k - t_1 & m_k - m_1 + d(t_k - t_1)
        \end{pmatrix},
        \quad
        2 \le j < k \le p.
    \]
    In the above, 
        $\det(A_{jk}) = 0$
    means 
        $\det(A_{jk}) \equiv 0 \mod p$, 
    and this notation is used for what follows.
    By simple linear algebra and \eqref{eq:a_i}, we have
    \[
        \det(A_{jk}) = 0 
        \Longleftrightarrow
        \det(A'_{jk}) = 0,
        \quad
        \text{ where matrix }
        A'_{jk} =
        \begin{pmatrix}
            n_j - n_1 & m_j - m_1 \\
            n_k - n_1 & m_k - m_1
        \end{pmatrix},
        \quad
        2 \le j < k \le p.
    \]
    Hence, it suffices to show 
        $\det(A'_{jk}) = 0$, 
    for some
        $2 \le j < k \le p$.
    Observe further that since $m_i$ are all distinct, one has for every 
        $2 \le i \le p$, 
        $m_i - m_1 \in \{1,\ldots, p-1\}$
    and these $p-1$ numbers are all distinct.
    Similarly,
        $n_i - n_1 \in \{1,\ldots, p-1\}$
    are all distinct.
    Hence, without loss of generality, we may assume 
        $m_i - m_1 = i-1$
    and let 
        $n_i - n_1 = x_i \in \{1, \ldots, p-1\}$
    for $2 \le i \le p$. In this notation, 
    the problem is then reduced to show 
    \begin{equation}   \label{eq:reduced-to-show}
        \det(A'_{jk}) \equiv 0 \mod p, 
        \quad
        \text{ for some } 
        2 \le j < k \le p,
    \end{equation}
    where
    \[
        A'_{jk} =
        \begin{pmatrix}
            x_j & j-1 \\
            x_k & k-1
        \end{pmatrix},
        \quad
        \text{ and }
        x_2, \ldots, x_p \in \{1, \ldots, p-1\}
        \text{ are distinct}.
    \]
    
    Above we have reduced the problem to  \eqref{eq:reduced-to-show} when the two directions are $(1,d)$ and $(1,d')$. The case when one of the two directions is $(0,1)$ can be handled in the same manner and can be reduced to \eqref{eq:reduced-to-show} too, we omit the detail.
    
    Finally, observe that \eqref{eq:reduced-to-show} is equivalent to Lemma \ref{lem:identity}. This completes the proof.
\end{proof}

\begin{proof}[Proof of Theorem \ref{thm:MBS}-(5)]
    Let 
        $A = \{a_1, \ldots, a_{p-1}\}$.
    Observe that the set
        $S = \ell \cup A$
    is a blocking set. Indeed, let $\ell'$ be a non-vertical line, then it either intersect $\ell$ or is parallel to $\ell$, in the latter case it passes through some $a_i$. Hence, $\ell' \cap S \neq \emptyset$ always hold.
    
    To show $S$ is an MBS, we say a point $x \in S$ is \emph{essential} if there is a non-vertical line $\ell^*$ passing through $x$ and $\ell^*$ is disjoint from $S \backslash \{x\}$. It suffices to show every point in $S$ is essential. 
    \begin{itemize}
        \item Every $a_i$ is essential: since there are $p-1$ lines parallel to $\ell$, each parallel line of $\ell$ passes through exactly one point $a_i$.
        
        \item As no two of $a_i$ are in the same vertical line, for $1\le i\le p-1$, let $\ell_i$ denote the unique vertical line passing through $a_i$, and let
            $b_i = \ell \cap \ell_i$.
        Every $b_i \in \ell$ is also essential. To see this, let 
            $\cL$
        be the set of the $p-1$ lines passing through $b_i$ that are not $\ell$. Let $\ell' \in \cL$, then
            $\ell' \cap (S \backslash \{b_i\})
            = \ell' \cap (A \backslash \{a_i\})$.
        And the intersection
            $\ell' \cap (A \backslash \{a_i\})$
        are disjoint from each for distinct $\ell' \in \cL$. As 
            $A \backslash \{a_i\} = p-2$,
        but $\cL$ contains $p-1$ lines, hence there must be a line $\ell^* \in \cL$ such that
            $\ell^* \cap (A \backslash \{a_i\}) = \emptyset$
        as desired.
        
        \item Since there are $p-1$ points $b_i$, there is one unique point $b \in \ell$ left. We show $b$ is also essential. Observe that the set of $p$ points 
            $\{a_1, \ldots, a_{p-1}, b\}$
        satisfies the condition of Lemma \ref{lem:intersecting-points-property}. Indeed, they are the $p$ intersecting points of the two sets of parallel lines: the set of $p$ vertical lines and the set of $p$ lines in the direction of $\ell$. Apply Lemma \ref{lem:intersecting-points-property}, there are $a_i, a_j$ such that $b, a_i, a_j$ are colinear. Repeating a similar argument as the previous case finishes the proof. \qedhere
    \end{itemize}
\end{proof}

\section{A read-k non-deterministic BP lower bound for GEN}   \label{sec:GEN}

In this section we derive an exponential lower bound for $\nBP_k$ for the GEN function via a direct reduction from the $\BRS$ function.

\begin{definition}  \label{def:proj}
A Boolean function\footnote{The domain is $\{0,1\}^n$ and the range is $\{0,1\}$.} sequence $f=(f_n)_{n\geq 1}$ is said to be a \emph{projection} of Boolean function $g=(g_n)_{n\geq 1}$, written $f \le_{p(n)} g$, if $f(x_1,\ldots,x_n)=g_{p(n)}(y_1,\ldots,y_{p(n)})$ for some polynomial $p$ and $y_j\in\{0,1,x_1,\overline{x_1},\ldots,x_n,\overline{x_n}\}$. If for every $i=1,\ldots, n$, the number of $j$ for which $y_j\in\{x_i,\overline{x_i}\}$ is at most $t$, then we denote the projection by  $f \le^{t}_{p(n)} g$.  
\end{definition}

\begin{lemma}  \label{lem:reduction}
    If 
        $f \le^{t}_{O(n^s)}  g$, 
    then 
        $\nBP_k(g_n) \ge \nBP_{kt}(f_{\Omega(n^{1/s})})$.
\end{lemma}

\begin{proof}
    Let $P$ be a nondeterministic BP for $g$. By the definition of the projection 
        $f \le^{t}_{O(n^s)}  g$, 
    replacing the variables in $P$ using variables for $f$ (or constants) gives a nondeterministic BP $Q$ for $f$. Hence, $|Q| \le |P|$.  Furthermore, $Q$ will be read-$(kt)$ if $P$ is read-$k$. This implies 
        $\nBP_{kt}(f_{n}) \le \nBP_k(g_{O(n^s)})$.
    Rewriting this inequality gives the lemma.
\end{proof}

\begin{remark}
By the construction in Proposition \ref{prop:small-S-large-NBP} one can show that, however, even $f \le^{1}_{O(n^s)}  g$ (i.e., so-called read-once projection) does not imply a similar inequality as in the Observation for the measures $\widehat{S}, S, \widehat{C}$ and $C$. That is, for example, one can have $f \le^{1}_{O(n^s)}  g$ and $\widehat{S}(f)$ is exponentially large but $\widehat{S}(g)$ is a constant. 
\end{remark}

\begin{theorem}  \label{thm:projection-to-GEN}
Let $f=(f_n)_{n\geq 1}$ be a sequence of Boolean functions so that $f_n$ has a De Morgan circuit of size $S(n) \ge n$ with the following property: no two gates have the same pair of gates as their two inputs. Then, $f \le^1_{O(S(n)^2 \log S(n))} \GEN$.
\end{theorem}
		
\begin{proof}
Fix a circuit $\cC_n$ for $f_n$ with the desired property, we define the projection from $\GEN$ as follows. Consider a set $M$ that contains the following elements:
\begin{itemize}
\item $n+1$ elements  $\$_0, \$_1, \ldots, \$_{n}$;
\item $2$ elements $(X_i,0), (X_i,1)$ for every $1 \le i \le n$;
\item $2$ elements $(g,0), (g,1)$ for every gate $g$ in the circuit $\cC_n$. 
\end{itemize} 
Let $m = |M|$, then $m = n+1+ 2n +2S(n) =  O(S(n))$. To compare this with the usual definition of $\GEN$ (see definition of $\GEN_q$ in Section \ref{sec:funcs}), we think of $\$_0$ as the element $1$, and $(g,1)$ as the $m$-th element in $M$, where $g$ is the output gate of $\cC_n$. Let $q=\frac{m(m-1)}{2} = O(S(n)^2)$. We show that $f_n$ is a projection of $\GEN_q: M^q \to \{0,1\}$. Specifically, given an input $n$-bit string $(x_1, \ldots, x_n) \in \{0,1\}^n$ for $f_n$, define the corresponding input $Y \in M^q$ for $\GEN_q$ as follows. 
\begin{enumerate}[(1)]
\item $\$_0 * \$_i = \$_{i+1}$ for every $0 \le i \le n-1$;
\item $\$_i * \$_i = (X_i,x_i)$ for every $1 \le i \le n$;
\item If $g$ is an $\land$ gate in $\cC_n$ with two input gates $h, l$, then define $(h,a)*(l,b) = (g,a \land b)$ for every $a,b\in \{0,1\}$, Define similarly if $g$ is an $\lor$ gate. 
\item If $g$ is a $\lnot$ gate with input gate $h$, then define $(h,0)*(h,0) = (g,1)$ and $(h,1)*(h,1) = (g,0)$.
\end{enumerate} 
Define $\delta * \gamma = \$_0$ for every $\delta, \gamma \in M$ where the operation $\delta * \gamma$ has not been defined above. Let the obtained variable for $\GEN_q$ be $Y \in M^q$. 
One can directly check that $f_n(x_1, \ldots, x_n) = \GEN_q(Y)$. Furthermore, by (2), each variable $x_i$ only appears once in $Y$, this proves the theorem.  
\end{proof}

\begin{corollary}  \label{cor:non-det-GEN}
$\nBP_k(\GEN_n) = 2^{\Omega(\frac{\sqrt{n}}{4^k k^3 \sqrt{\log n}})}$.
\end{corollary}

\begin{proof}

It is not hard to construct a size $O(n)$ De Morgan circuit for $\BRS_n$ satisfying the property in Theorem \ref{thm:projection-to-GEN}. Hence, $\BRS \le^1_{O(n^2\log n)} \GEN$. In \cite{BRS1993} it is shown that $\nBP_k(\BRS_n) = 2^{\Omega(\frac{n}{4^k k^3})}$.  Lemma \ref{lem:reduction} implies the desired lower bound for $\GEN$. 
\end{proof}

\section{Discussion and open problems}   \label{sec:openproblem}

Although exponential lower bounds for deterministic read-once BPs have been proved more than three decades ago \cite{vzak1984exponential,wegener1988complexity}, the read-once BP models still offer challenges such as (a) finding a function having small DNF and small CNF, yet having no small deterministic read-once BP, and (b) finding a Boolean function having an exponential lower bound for \emph{semantic} nondeterministic read-once BP (we did not discuss this topic in our paper, see reference \cite{cook2016lower}). Besides, read-once BPs are also important for their connections to derandomization and proof complexity. Below we discuss questions naturally inspired  from our work. 

\begin{enumerate}[(1)]
    \item What is $\widehat{P}(\TEP,h)$? By Theorem \ref{thm:hatP-BP1}, an exponential lower bound for $\widehat{P}(\TEP,h)$ would give an alternative proof, besides \cite{iwama2018read}, for $(\TEP,h)$ having exponential deterministic  $\BPone$  lower bound. The reason to pursue an alternative proof is the hope that it might be generalized to read-$k$ BP. Indeed, $\widehat{\cP_k}$ as a natural generalization of $\widehat{P}$ is a lower bound for $\BP_k$ as discussed in Section \ref{sec:read-k}. Similarly, what is $\widehat{C}(\TEP,h)$? As well, we have not been able to solve $S(\TEP,h)$, despite the fact that the only immediate implication of an $S$ lower bound is a lower bound on the size of OBDDs, a very weak BP model. 
    
    \item What is $\CC(\TEP,h)$? An easy upper bound is $\CC(\TEP,h) = O(h \log k)$ following from Theorem \ref{thm:S-TEP}. Referring to Figure \ref{fig:relation}, note that a lower bound for $\CC(\TEP,h)$ does not imply lower bounds for $\widehat{P}(\TEP,h)$ or $\widehat{C}(\TEP,h)$, or vice versa. Nonetheless, studying the communication complexity of TEP seems to be an interesting question itself. 
    
    \item In view of Theorem \ref{thm:Gal-mixed}, does there exist an  $r$-regular and $K_{2,s}$-free  bipartite graph $G$, such that $\widehat{S}(\GAL_G)$ is large but $\widehat{C}(\GAL_G)$ is small, for appropriate parameters $r$ and $s$? This means that for the purpose of tackling the  $\BPone$ vs weight problem there is no reason to restrict oneself to the graph $G_q$. Of course, determining $\widehat{C}(\GAL_{G_q})$ is still an interesting problem for $\nBPone(\GAL_{G_q})$.
    
    \item The {\bf Question 1} and {\bf Question 2} asked in Section \ref{sec:Gal-BW-SChat}. Question 2 is a typical question in studying the standard blocking sets in either projective or affine spaces. In contrast, the number of blocking sets  satisfying certain conditions (i.e., Question 1) receives little attention (see \cite{hirschfeld1998projective,szHonyi2003spectrum}). Our work provides a strong motivation for it. Yet, to fully classify blocking sets in $\F_p^2$ in order to help answer Question 1 seems an daunting task. Is it possible to solve Question 1 without answering Question 2, and in general, is it possible to determine $\widehat{C}(\GAL_{G_q})$ without answering  Question 2? This again points to the direction that, if one is only interested in complexity applications, perhaps it will be useful to try other graphs besides $G_q$. Having said that, studying the number and structures of (minimal) blocking sets in $\F_p^2$ are interesting mathematical problems on their own, as we try to demonstrate in Section \ref{sec:minimal-BS}.
    
    \item By Proposition \ref{prop:small-S-large-NBP} and Figure \ref{fig:relation}, none of the measures (such as $S$, $\widehat{C}$, etc) defined in the paper lie in-between $\BPone$ and circuit size. In view of derandomization, it could be beneficial to have a sequence of measures, say $m_1, m_2, \ldots$, such that 
        $\BPone \ge m_1 \ge m_2 \ge \ldots \ge$
    circuit size. This in theory might allow a progressive way of adapting techniques for derandomizing space (i.e., $\BPone$), where things are better understood, to derandomizing time (i.e., circuit size). For example, a typical question could be to construct pseudo random generators for Boolean functions $f$ satisfying $m_i(f) \le O(n^2)$.
\end{enumerate}

\section*{Acknowledgement}
The authors acknowledge support from the Natural Sciences and Engineering Research Council of Canada, P.M.\ as holder of discovery grant RGPIN-04500 and Y.L.\ as postdoctoral collaborator. 
Y.L. acknowledges the support from University of Montr{\'e}al where this work was done when Y.L. was a postdoctoral researcher there.

\bibliography{mybib}{}

\begin{thebibliography}{10}

\bibitem{NeciBoundTight2016}
Paul Beame, Nathan Grosshans, Pierre McKenzie, and Luc Segoufin.
\newblock Nondeterminism and an abstract formulation of {N}e\v{c}iporuk's lower
  bound method.
\newblock {\em ACM Trans. Comput. Theory}, 9(1):Art. 5, 34, 2016.

\bibitem{blokhuis2002combinatorial}
A~Blokhuis.
\newblock Combinatorial problems in finite geometry and lacunary polynomials.
\newblock In {\em Proceedings International Congress of Mathematicians (ICM
  2002, Beijing, China, August 20-28, 2002), Volume III: Invited lectures},
  pages 537--545. Higher Education Press, 2002.

\bibitem{bollig1998very}
Beate Bollig and Ingo Wegener.
\newblock A very simple function that requires exponential size read-once
  branching programs.
\newblock {\em Information Processing Letters}, 66(2):53--57, 1998.

\bibitem{BRS1993}
Allan Borodin, Alexander Razborov, and Roman Smolensky.
\newblock On lower bounds for read-k-times branching programs.
\newblock {\em Computational Complexity}, 3(1):1--18, 1993.

\bibitem{brouwer1978blocking}
Andries~E Brouwer and Alexander Schrijver.
\newblock The blocking number of an affine space.
\newblock {\em Journal of Combinatorial Theory, Series A}, 24(2):251--253,
  1978.

\bibitem{tashma2021PrgsAgainstReadOnce}
Gil Cohen, Dean Doron, Oren Renard, Ori Sberlo, and Amnon Ta{-}Shma.
\newblock Error reduction for weighted prgs against read once branching
  programs.
\newblock In Valentine Kabanets, editor, {\em 36th Computational Complexity
  Conference, {CCC} 2021, July 20-23, 2021, Toronto, Ontario, Canada (Virtual
  Conference)}, volume 200 of {\em LIPIcs}, pages 22:1--22:17. Schloss Dagstuhl
  - Leibniz-Zentrum f{\"{u}}r Informatik, 2021.

\bibitem{cook2016lower}
Stephen Cook, Jeff Edmonds, Venkatesh Medabalimi, and Toniann Pitassi.
\newblock Lower bounds for nondeterministic semantic read-once branching
  programs.
\newblock In {\em 43rd International Colloquium on Automata, Languages, and
  Programming (ICALP 2016)}. Schloss Dagstuhl-Leibniz-Zentrum fuer Informatik,
  2016.

\bibitem{cook2012pebbles}
Stephen Cook, Pierre McKenzie, Dustin Wehr, Mark Braverman, and Rahul
  Santhanam.
\newblock Pebbles and branching programs for tree evaluation.
\newblock {\em ACM Transactions on Computation Theory (TOCT)}, 3(2):1--43,
  2012.

\bibitem{ehrenfeucht1989learning}
Andrzej Ehrenfeucht and David Haussler.
\newblock Learning decision trees from random examples.
\newblock {\em Information and Computation}, 82(3):231--246, 1989.

\bibitem{forbes2018pseudorandom}
Michael~A Forbes and Zander Kelley.
\newblock Pseudorandom generators for read-once branching programs, in any
  order.
\newblock In {\em 2018 IEEE 59th Annual Symposium on Foundations of Computer
  Science (FOCS)}, pages 946--955. IEEE, 2018.

\bibitem{gal1997simple}
Anna G{\'a}l.
\newblock A simple function that requires exponential size read-once branching
  programs.
\newblock {\em Information Processing Letters}, 62(1):13--16, 1997.

\bibitem{Tseitin2017}
Ludmila Glinskih and Dmitry Itsykson.
\newblock Satisfiable tseitin formulas are hard for nondeterministic read-once
  branching programs.
\newblock In {\em 42nd International Symposium on Mathematical Foundations of
  Computer Science (MFCS 2017)}. Schloss Dagstuhl-Leibniz-Zentrum fuer
  Informatik, 2017.

\bibitem{Tseitin2019}
Ludmila Glinskih and Dmitry Itsykson.
\newblock On tseitin formulas, read-once branching programs and treewidth.
\newblock {\em Theory of Computing Systems}, pages 1--21, 2020.

\bibitem{hirschfeld1998projective}
JWP Hirschfeld.
\newblock {\em Projective geometries over finite fields. Oxford mathematical
  monographs}.
\newblock Oxford University Press New York, 1998.

\bibitem{TseitinOBDD2017}
Dmitry Itsykson, Alexander Knop, Andrei Romashchenko, and Dmitry Sokolov.
\newblock On obdd-based algorithms and proof systems that dynamically change
  order of variables.
\newblock {\em The Journal of Symbolic Logic}, pages 1--41, 2020.

\bibitem{iwama2018read}
Kazuo Iwama and Atsuki Nagao.
\newblock Read-once branching programs for tree evaluation problems.
\newblock {\em ACM Transactions on Computation Theory (TOCT)}, 11(1):1--12,
  2018.

\bibitem{JonesGEN}
Neil~D. Jones and William~T. Laaser.
\newblock Complete problems for deterministic polynomial time.
\newblock {\em Theor. Comput. Sci.}, 3(1):105--117, 1976.

\bibitem{jukna2012boolean}
Stasys Jukna.
\newblock {\em Boolean function complexity: advances and frontiers}, volume~27.
\newblock Springer Science \& Business Media, 2012.

\bibitem{jukna1999p}
Stasys Jukna, A~Razborov, P~Savicky, and Ingo Wegener.
\newblock On p versus np $\cap$ co-np for decision trees and read-once
  branching programs.
\newblock {\em Computational Complexity}, 8(4):357--370, 1999.

\bibitem{mixed}
Stasys~P Jukna.
\newblock Entropy of contact circuits and lower bounds on their complexity.
\newblock {\em Theoretical Computer Science}, 57(1):113--129, 1988.

\bibitem{kushilevitz1997communication}
Eyal Kushilevitz and Noam Nisan.
\newblock Communication complexity, 1997.

\bibitem{rao2020communication}
Anup Rao and Amir Yehudayoff.
\newblock {\em Communication Complexity: and Applications}.
\newblock Cambridge University Press, 2020.

\bibitem{razgon2014obdds}
Igor Razgon.
\newblock On obdds for cnfs of bounded treewidth.
\newblock In {\em Proceedings of the Fourteenth International Conference on
  Principles of Knowledge Representation and Reasoning}, pages 92--100, 2014.

\bibitem{sauerhoff2003approximation}
Martin Sauerhoff.
\newblock Approximation of boolean functions by combinatorial rectangles.
\newblock {\em Theoretical computer science}, 301(1-3):45--78, 2003.

\bibitem{SS1993}
Janos Simon and Mario Szegedy.
\newblock A new lower bound theorem for read-only-once branching programs and
  its applications.
\newblock In {\em Advances in Computational Complexity Theory}, pages 183--193,
  1990.

\bibitem{sofronovaSokolov2021boundedRepetition}
Anastasia Sofronova and Dmitry Sokolov.
\newblock Branching programs with bounded repetitions and flow formulas.
\newblock In Valentine Kabanets, editor, {\em 36th Computational Complexity
  Conference, {CCC} 2021, July 20-23, 2021, Toronto, Ontario, Canada (Virtual
  Conference)}, volume 200 of {\em LIPIcs}, pages 17:1--17:25. Schloss Dagstuhl
  - Leibniz-Zentrum f{\"{u}}r Informatik, 2021.

\bibitem{szHonyi2003spectrum}
Tam{\'a}s Sz{\H{o}}nyi, Andr{\'a}s G{\'a}cs, and Zsuzsa Weiner.
\newblock On the spectrum of minimal blocking sets in pg $(2, q) $.
\newblock {\em Journal of Geometry}, 76(1-2):256--281, 2003.

\bibitem{wegener1987complexity}
Ingo Wegener.
\newblock {\em The complexity of Boolean functions}.
\newblock BG Teubner, 1987.

\bibitem{wegener1988complexity}
Ingo Wegener.
\newblock On the complexity of branching programs and decision trees for clique
  functions.
\newblock {\em Journal of the ACM (JACM)}, 35(2):461--471, 1988.

\bibitem{wegener2000branching}
Ingo Wegener.
\newblock {\em Branching programs and binary decision diagrams: theory and
  applications}.
\newblock SIAM, 2000.

\bibitem{vzak1984exponential}
Stanislav {\v{Z}}{\'a}k.
\newblock An exponential lower bound for one-time-only branching programs.
\newblock In {\em International Symposium on Mathematical Foundations of
  Computer Science}, pages 562--566. Springer, 1984.

\end{thebibliography}
\bibliographystyle{plain}

\appendix

\section{The proof for Theorem \ref{thm:S-TEP}} \label{sec:appendix}

In this appendix, we provide calculation and proof for Theorem \ref{thm:S-TEP}. We will prove each part of Theorem \ref{thm:S-TEP} separately, as in Theorem \ref{thm:S-TEP2}, Theorem \ref{thm:S-TEP3} and Theorem \ref{thm:S-TEP-h}, in the following.

We use the following notation.  Let $(M, L, R)$ denote a partition of the input variables for  $(\TEP,h)$ where $M$ corresponds to the root matrix, $L$ and $R$ correspond to the left and right child, respectively. Note that both $L$ and $R$ correspond to inputs for $(\TEP,h-1)$. Recall $n_h$ denotes the input size for $(\TEP, h)$. For a subset $A \subseteq [n_h]$, we think of $A$ as a subset of input variables for $(\TEP,h)$ and write $A = (A_M, A_L, A_R)$, where $A_M = A \cap M$, $A_L = A \cap L$ and $A_R = A \cap R$. For notational simplicity, when the parameter $h$ is clear from the context, we use $\TEP_A$ to denote the matrix $(\TEP,h)_A$, and for $\alpha \in [k]^A$, we use $\TEP_\alpha$ to denote the subfunction $(\TEP,h)_{A,\alpha}$. 

By an abuse of notation, sometimes we use $M_{ij} \in A_M$ or $(i,j) \in A_M$, depending on which one is more convenient in the context, to mean that $A_M$ contains the variable at entry $(i,j)$ of the root matrix $M$.

Let $A \subseteq [n_h]$.
Let $\alpha, \alpha' \in [k]^{A}$. We say
    $\alpha \sim \alpha'$ 
if 
    $(\TEP,h)_{A,\alpha} = (\TEP,h)_{A,\alpha'}$.
Obviously, $\sim$ is an equivalence relation. 
We use the notation 
    $\langle \alpha \rangle$ 
to denote the equivalence class represented by $\alpha$.

A function is said to be \emph{non-constant} if it evaluates to at least two distinct values. 
A subfunction $(\TEP,h)_{A,\alpha}$ is a \emph{full-range} function if 
for every 
    $r \in [k]$
there exists 
    $\beta \in [k]^{[n_h] \setminus A}$
satisfying
	$(\TEP,h)_{A,\alpha}(\beta) = r$.

\begin{lemma}   \label{lem:non-constant-subfunc}
	Let $h\ge 1$.
	For any 
	    $A \subseteq [n_h]$ of size
	    $|A| \le n_h - 1$,
	there exists
	    $\alpha^* \in [k]^{A}$
	such that $(\TEP,h)_{A,\alpha^*}$ is a full-range function.
\end{lemma}

\begin{proof}
    We use induction on $h$. The base case $h=1$ is clear. 
    Assume the lemma is true for $h-1$. 
    Consider $(\TEP,h)$. Let $A=(A_M, A_L, A_R)$ satisfy $|A| \le n_h - 1$.
    \begin{itemize}
        \item $A_L \neq L$, i.e., $|A_L| \le n_{h-1} - 1$.
            By induction hypothesis on $h-1$, there exists 
                $\alpha^*_L \in [k]^{A_L}$
            such that 
                $(\TEP,h-1)_{A_L,\alpha^*_L}$
            is a full-range function.
            Fix 
                $\alpha^*_R \in [k]^{A_R}$
            and 
                $\beta_R \in [k]^{R-A_R}$
            arbitrarily. Suppose
                $(\TEP,h-1)_{A_R, \alpha^*_R}(\beta_R) = j$
            for some $j \in [k]$.
            Fix 
                $\alpha^*_M \in [k]^{A_M}$
            and 
                $\beta_M \in [k]^{M-A_M}$ 
            such that the $j$-th column of the matrix $M$ is the vector
                $(1,2,..., k)^T$.
            For every $r \in [k]$, there exists
                $\beta_L \in [k]^{L-A_L}$
            such that
                $(\TEP,h-1)_{A_L,\alpha^*_L}(\beta_L) = r$.
            Set 
                $\alpha^* = (\alpha^*_M, \alpha^*_L, \alpha^*_R)$
            and
                $\beta = (\beta_M, \beta_L, \beta_R)$.
            Then, 
                $(\TEP,h)_{A,\alpha^*}(\beta) = r$.
                
        \item $A_R \neq R$. This is symmetric to the previous case.
        
        \item $A_L = L$ and $A_R = R$ but $A_M \neq M$.
            Choose $\alpha^*_L$ and $\alpha^*_R$
            such that 
                $\big( (\TEP,h-1)(\alpha^*_L), (\TEP,h-1)(\alpha^*_R) \big) = (i,j) \not\in A_M$.
            Set $\alpha^*_M$ arbitrarily. 
            Let
                $\alpha^* = (\alpha^*_M, \alpha^*_L, \alpha^*_R)$.
            It is easy to see that 
                $(\TEP,h)_{A,\alpha^*}$
            is a full-range function.  \qedhere
    \end{itemize}
\end{proof}

\begin{lemma}   \label{lem:equivalence}
Let 
	$h \ge 2$, 
	$A\subseteq [n_h]$. 
Suppose 
	$A=(\emptyset,A_L,A_R)$.  
Let 
	$\alpha= (\alpha_L,\alpha_R) \in [k]^{A} = [k]^{A_L} \times [k]^{A_R}$, 
	$\alpha' = (\alpha'_L,\alpha'_R) \in [k]^{A} = [k]^{A_L} \times [k]^{A_R}$. 
Then,
	\begin{align}  \label{eq:equivalence}
		&(\TEP,h)_{A,\alpha} = (\TEP,h)_{A,\alpha'}  \nonumber \\
		&\Longleftrightarrow
		\Big( (\TEP,h-1)_{A_L,\alpha_L}, (\TEP,h-1)_{A_R,\alpha_R} \Big)  = 
		\Big( (\TEP,h-1)_{A_L,\alpha'_L}, (\TEP,h-1)_{A_R,\alpha'_R} \Big).
	\end{align} 
\end{lemma}

\begin{proof}
The direction $\Longleftarrow$. Obvious.

The direction $\Longrightarrow$. 
	Assume for the sake of a contradiction the implication is not true. 
	Without loss of generality we may assume 
		$(\TEP,h-1)_{A_L,\alpha_L} \neq  (\TEP,h-1)_{A_L,\alpha'_L}$. 
	We will show
	\[
		(\TEP,h-1)_{A_L,\alpha_L} \neq  (\TEP,h-1)_{A_L,\alpha'_L}
		\Longrightarrow (\TEP,h)_{A,\alpha} \neq (\TEP,h)_{A,\alpha'}.
	\]
	Indeed, let 
		$\beta_L \in [k]^{L-A_L}$ 
		be such that 
		    $(\TEP,h-1)_{A_L,\alpha_L}(\beta_L) = i
		    \neq 
		    i' = (\TEP,h-1)_{A_L,\alpha'_L}(\gamma_L)$.
	Then, set $\beta_M \in [k]^{M}$ to be such that every entry in the $i$-th row equals to $i$, and every entry in the $i'$-th row equals to $i'$. 
	Choose $\beta_R  \in [k]^{R-A_R}$ arbitrarily. 
	Set
	    $\beta = (\beta_M,\beta_L,\beta_R)$.
	Then, 
	    $(\TEP,h)_{A,\alpha}(\beta) = i
	    \neq i'
	    = (\TEP,h)_{A,\alpha'}(\beta)$.
\end{proof}

\begin{lemma}   \label{lem:easy-case}
	Let $h \ge 2$. 
	Let 
		$A=(A_M,A_L,A_R) \subseteq [n_h]$.
	If 
		$|A_L|, |A_R| \le n_{h-1} - 1$,
	then
		$S\big( (\TEP,h)_A \big) \ge k^{|A_M|}$.
\end{lemma}

\begin{proof}
	Since 
		$|A_L|\le n_{h-1} - 1$,
	Lemma \ref{lem:non-constant-subfunc} implies that there exists $\alpha^*_L \in [k]^{A_L}$ such that
		$(\TEP,h-1)_{A_L, \alpha^*_L}$ 
	is a full-range function. Similarly, let $\alpha^*_R \in [k]^{A_R}$ be such that 
	    $(\TEP,h-1)_{A_R, \alpha^*_R}$ 
	is a full-range function. 
	Consider the set 
		\[
			\Omega = \{\alpha = (\alpha_M, \alpha^*_L, \alpha^*_R): \alpha_M \in [k]^{A_M}\}.
		\]
	We claim that 
		\[
			\alpha, \alpha' \in \Omega, 
			\alpha \neq \alpha'
			\Longrightarrow
			(\TEP,h)_{A,\alpha} \neq (\TEP,h)_{A,\alpha'}.
		\]
	This implies 
		$S\big( (\TEP,h)_A \big) \ge |\Omega| = k^{|A_M|}$.

	To show the claim, suppose $\alpha_M(i,j) \neq \alpha'_M(i,j)$ for some $(i,j) \in A_M$. 
	By the choice of $\alpha^*_L$ and $\alpha^*_R$, there exist 
		$(\beta_L, \beta_R) \in [k]^{L - A_L} \times [k]^{R- A_R}$
	such that
		$(\TEP,h-1)_{A_L,\alpha^*_L}(\beta_L) = i$ and
		$(\TEP,h-1)_{A_L,\alpha^*_R}(\beta_R) = j$.
	Let $\beta = (\beta_M, \beta_L, \beta_R)$ where $\beta_M \in [k]^{M-A_M}$ is chosen arbitrarily. Then
		\[
			(\TEP,h)_{A,\alpha}(\beta) = \alpha_M(i,j)
			\neq 
			\alpha'_M(i,j) = (\TEP,h)_{A,\alpha'}(\beta)
		\]
	as claimed.
\end{proof}

\begin{lemma}   \label{lem:one-row}
	Let $h \ge 2$. Let 
		$A=(A_M,L,A_R)$,
	i.e., 
		$A_L = L$.
	Suppose
		$|A_R| \le n_{h-1} - 1$. 
	For $i\in [k]$, let
	    $r_i$
	denote the number of entries of $A_M$ in row $i$.
	\begin{enumerate}[(1)]
	    \item If $r_i=k$ for some $i$, then
	        $S((\TEP,h)_A) \ge k^k$.
	        
	    \item  If 
	            $r_i < k$
        	for every $i \in [k]$, then
	            $S((\TEP,h)_A) \ge \sum_{i=1}^k k^{r_i}$.
	\end{enumerate}
\end{lemma}

\begin{proof}
    For $i\in [k]$, let 
	    $\langle i \rangle$
	denote 
	    $\langle \alpha_L \rangle$
	for which 
	    $(\TEP,h-1)(\alpha_L) = i$. 
	Lemma \ref{lem:non-constant-subfunc} implies the existence of 
	    $\alpha^*_R \in [k]^{A_R}$ 
	such that
		$(\TEP,h-1)_{A_R, \alpha^*_R}$ is a full-range function.
    
    {\bf Claim:} There are exactly
        $k^{r_i}$
    distinct subfunctions $(\TEP,h)_{A,\alpha}$ for which $\alpha$ is of the form
        $\alpha = (\alpha_M, \langle i \rangle,\alpha^*_R) \in [k]^{A}$.
        
    Proof of the Claim: Let 
		$\alpha' = (\alpha'_M, \langle i \rangle,\alpha^*_R) \in [k]^{A}$. 
	Suppose $\alpha_M$ and $\alpha'_M$  differ on row $i$, i.e.,	
		$\alpha_M(i, j) \neq \alpha'_M(i,j)$
	for some $j$ where $(i,j) \in A_M$. 
	It suffices to show $(\TEP,h)_{A,\alpha} \neq (\TEP,h)_{A,\alpha'}$.
	Indeed, since $(\TEP,h-1)_{A_R,\alpha^*_R}$ is a full-range function, there exists 
		$\beta_R \in [k]^{R-A_R}$
	such that
		$(\TEP,h-1)_{A_R,\alpha^*_R}(\beta_R) = j$.
	Let 
		$\beta = (\beta_M, \beta_R) \in [k]^{A-A_M} \times  [k]^{R-A_R}$
	where $\beta_M$ is chosen arbitrarily.
	Then,
		\[
			(\TEP,h)_{A,\alpha}(\beta) = \alpha_M(i, j) 
			\neq 
			\alpha'_M(i,j) = (\TEP,h)_{A,\alpha'}(\beta).
		\]
	Hence, 
		$(\TEP,h)_{A,\alpha} \neq (\TEP,h)_{A,\alpha'}$.
	Since the entries in $A_M$ that are not in row $i$ are irrelevant for the subfunction $(\TEP,h)_{A,\alpha}$ for which $\alpha$ is of the form defined before, the number of such subfunctions is equal to $k^t$. 
	
	We proceed to prove the Lemma.
	\begin{enumerate}[(1)]
	    \item This follows directly from the Claim.
	    
	    \item By the Claim, it suffices to show subfunctions given by
        	    $\alpha = (\alpha_M, \langle i \rangle, \alpha^*_R)$
        	are distinct for distinct $i \in [k]$.
        	
            Let $i,i' \in [k]$ and $i\neq i'$. Consider
                $\langle i \rangle$
            and 
                $\langle i' \rangle$.
            Let
                $\alpha = (\alpha_M, \langle i \rangle, \alpha^*_R)$
            and
                $\alpha' = (\alpha'_M, \langle i' \rangle, \alpha^*_R)$.
            We show 
                $(\TEP,h)_{A,\alpha} \neq (\TEP,h)_{A,\alpha'}$.
            Since 
                $r_i < k$,
            there exists $j \in [k]$ such that
                $(i,j) \not\in A_M$.
            Choose $\beta_R$ as before such that 
                $(\TEP,h-1)_{A_R,\alpha^*_R}(\beta_R) = j$.    
            \begin{itemize}
                \item Case 1: $(i',j) \not\in A_M$. 
                    Choose 
                        $\beta_M \in [k]^{M - A_M}$ 
                    such that
                        $\beta_M(i,j) = 1$
                    and
                        $\beta_M(i',j) = 2$.
                    Set $\beta= (\beta_M, \beta_R)$. Then, 
                        $(\TEP,h)_{A,\alpha}(\beta) = 1$ 
                    but 
                        $(\TEP,h)_{A,\alpha'}(\beta) = 2$.
                \item Case 2: $(i',j) \in A_M$. 
                    Choose 
                        $\beta_M \in [k]^{M - A_M}$ 
                    such that
                        $\beta_M(i,j) \neq \alpha'_M(i',j)$.
                    Set $\beta= (\beta_M, \beta_R)$. Then, 
                        $(\TEP,h)_{A,\alpha}(\beta) = \beta_M(i,j)
                            \neq 
                            \alpha'_M(i',j) = (\TEP,h)_{A,\alpha'}(\beta). 
                        $
            \end{itemize}   
            In both cases, we have 
                 $(\TEP,h)_{A,\alpha} \neq (\TEP,h)_{A,\alpha'}$
            as desired.         \qedhere
	\end{enumerate}
\end{proof}

\begin{lemma}   \label{lem:A_M-small}
	Let $h \ge 2$. Let 
		$A=(A_M,A_L,A_R)$.
	If $|A_M| \le k$, then
	    $S((\TEP,h)_{A}) \ge S((\TEP,h-1)_{A_L}) \cdot S((\TEP,h-1)_{A_R})$.
	Furthermore, if $|A_M| = 0$, then the equality holds.
\end{lemma}

\begin{proof}

  The ``Furthermore'' part follows from Lemma \ref{lem:equivalence}.
    
    To show the inequality, choose 
        $\alpha^*_M \in [k]^{A_M}$ 
    such that $\alpha^*_M$ assigns distinct values for entries in $A_M$. This is possible because 
        $|A_M| \le k$.
    Let 
        $(\alpha_L, \alpha_R) \in [k]^{A_L} \times [k]^{A_R}$.
    It suffices to show that each different pair
        $(\langle \alpha_L \rangle, \langle \alpha_R \rangle)$
    gives rise to a different subfunction
        $(\TEP,h)_{A,\alpha}$
    where 
        $\alpha = (\alpha^*_M,\alpha_L, \alpha_R)$.
    Note that the number of distinct pairs 
        $(\langle \alpha_L \rangle, \langle \alpha_R \rangle)$
    is exactly the desired lower bound.
    
    To verify the claim, consider two distinct pairs
        \begin{equation}  \label{eq:distinct}
            (\langle \alpha_L \rangle, \langle \alpha_R \rangle)
            \neq 
            (\langle \alpha'_L \rangle, \langle \alpha'_R \rangle).
        \end{equation}
    Let 
        $\alpha = (\alpha^*_M, \langle \alpha_L \rangle, \langle \alpha_R \rangle)$ 
    and 
        $\alpha' = (\alpha^*_M, \langle \alpha'_L \rangle, \langle \alpha'_R \rangle)$.
    The assumption \eqref{eq:distinct} implies that there exists 
        $(\beta_L, \beta_R) \in [k]^{L- A_L} \times [k]^{R- A_R}$
    such that
        \begin{align*}
            &\big( (\TEP,h-1)_{A_L,\alpha_L}(\beta_L), (\TEP,h-1)_{A_R,\alpha_R}(\beta_R) \big) \\
            &= (i,j)
            \neq 
            (i',j') 
            = \big( (\TEP,h-1)_{A_L,\alpha'_L}(\beta_L), (\TEP,h-1)_{A_R,\alpha'_R}(\beta_R) \big).
        \end{align*}
    By the choice of  $\alpha^*_M$, it is easy to see that there exists 
        $\beta_M \in [k]^{M - A_M}$
    such that
        $(\TEP,h)_{A,\alpha}(\beta) \neq (\TEP,h)_{A,\alpha'}(\beta)$
    for 
        $\beta = (\beta_M, \beta_L, \beta_R)$.
\end{proof}

Let 
    $S((\TEP,h), \ell) = \min_{A \subseteq [n_h],|A| = \ell} S((\TEP,h)_A)$.
Then,
    $S(\TEP,h) = \max_{1 \le \ell \le n_h} S((\TEP,h), \ell)$.

\begin{theorem}  \label{thm:S-TEP2}
    Let $1 \le \ell \le n_2$ be an integer.
    Let $c \in [0,1)$. Define
    $V_c = k+2$ and 
    $W_c = ck^2$.
    \begin{enumerate}[(1)]
        \item For $1 \le \ell \le k+1$,
            $S((\TEP,2), \ell) = S\big( (\TEP,2), n_2 - (k-1)(\ell-1) \big)$.
        
        \item  Let $c \in [0,1)$. If
                $V_c \le \ell \le W_c$,
            then
                $S\big( (\TEP,2),\ell \big) \ge (1-c)k^2$.
            In particular,
                \[
                    S(\TEP,2) 
                    = \max_{1 \le \ell \le k+1} S((\TEP,2), \ell)
                    = \max_{k+2 \le \ell \le n_2} S((\TEP,2), \ell) = k^2,
                \]
            and is  achieved at either $\ell = k+1$ or $\ell = k+2$.
    \end{enumerate}
\end{theorem}  

\begin{proof}
	Recall $n_2 = k^2 + 2$. 
	Assume the calculation of $S\big( (\TEP,2)_A \big)$ in Table \ref{table:S-TEP2} is correct.
	For every 	$1 \le \ell \le n_2$, in Table \ref{table:S-TEP2-minimizer} we give a minimizer 
		$A^* \subseteq [n_2]$ 
	of size 
		$|A^*| = \ell$
	such that
		$S((\TEP,2)_{A^*}) = S( (\TEP,2), \ell )$.
	The minimizer is obtained from Table \ref{table:S-TEP2}.
	The theorem follows by a simple calculation using Table \ref{table:S-TEP2-minimizer}. 
	
	We now prove the calculation of $S\big( (\TEP,2)_A \big)$ in Table \ref{table:S-TEP2} is correct. Let $|A| = \ell$. 
	There are four cases.
	\begin{itemize}
		\item 	$A=(A_M,\emptyset,\emptyset)$. 
				The upper bound is trivial, the lower bound follows from Lemma \ref{lem:easy-case}.
				
		\item  	$A=(A_M,x,\emptyset)$. 
				
				Consider the case 
					$|A_M| = \ell - 1 > (k-1)k$
				first. This implies that $A_M$ must contain all the $k$ entries for some row.
				Then, Lemma \ref{lem:one-row} implies 
					$S\big( (\TEP,2)_A \big) \ge k^k$
				
				Now assume $|A_M| = \ell - 1 \le (k-1)k$. 
				Observe that this implies  
					$qk^{p+1} +(k-q)k^p \le k^k$
				where 
					$\ell-1 = pk+q$.
					
				We consider two cases depending on the choice of $A_M$.
				\begin{itemize}
					\item Case 1: $A_M$ contains all the $k$ entries of some row. 
							By Lemma \ref{lem:one-row},
								$S\big( (\TEP,2)_A \big) \ge k^k$.	
					
					\item Case 2: $A_M$ contains at most $k-1$ entries from each row.
							Then, Lemma \ref{lem:one-row} implies that to minimize $S\big( (\TEP,2)_A \big)$
							the $A_M$ should be chosen greedily according to the columns, as described in Table \ref{table:S-TEP2}. For this choice of $A_M$, Lemma \ref{lem:one-row} implies 
								$S\big( (\TEP,2)_A \big) = qk^{p+1} +(k-q)k^p \le k^k$.
				\end{itemize}
				The above two cases imply that the optimal choice for $A_M$ is as described in Table \ref{table:S-TEP2}.

		\item  $A=(\emptyset,x,y)$. Obvious.
		
		\item  $A=(A_M,x,y)$. 
				Let 
					$\alpha = (\alpha_M, i, j) \in [k]^{A}$.
				\begin{itemize}
					\item	Case 1: $(i, j) \in A_M$.
							In this case, for all 
								$\gamma \in [k]^{\overline{A}}$, 
							one has
								\begin{equation}   \label{eq:eval-to-const-func}
									(\TEP,2)_{A,\alpha}(\gamma) = \alpha_M(i, j),
								\end{equation} 
							i.e., $(\TEP,2)_{A,\alpha}$ is a constant function that is identically equal to  $\alpha_M(i, j)$. Trivially, there are $k$ distinct constant functions. Note that
									$A_M \neq \emptyset$
								because 
									$\ell \ge 3$, 
								Hence, by \eqref{eq:eval-to-const-func}, each constant function can be achieved by some $\alpha \in [k]^{A}$ .
							 
					\item	Case 2: $(i, j) \not\in A_M$. 
							Let 
								$\gamma \in [k]^{\overline{A}}$.
							Then,
								$(\TEP,2)_{A,\alpha}(\gamma) = \gamma(i, j)$.
							Let
								$\alpha' = (\alpha'_M, i',j') \in [k]^{A}$
							such that
								$(i', j') \not\in A_M$.
							Then, it is easy to see that								
								$(\TEP,2)_{A,\alpha} = (\TEP,2)_{A,\alpha'}$ 
							is equivalent to 
								$(i, j) = (i', j')$.
							In other words,  every pair $(i, j) \not\in A_M$ defines a distinct subfunction, and no more. 
							Hence, there are $k^2 - |A_M|$ such subfunctions.
				\end{itemize}
				
				To summarize the two cases, we get that
					$S\big( (\TEP,2)_A \big) = k + (k^2 - |A_M|) = k^2 + k + 2 - \ell$.
	\end{itemize}
	Note that the case $A=(A_M,\emptyset, y)$ is symmetric to the case $A=(A_M,x,\emptyset)$. Hence, we have verified all possible cases. 
\end{proof}

\begin{table}[ht!]
	\centering
	\caption{The calculation for $S\big( (\TEP,2)_A \big)$, categorized according to the pattern of $A=(A_M,A_L,A_R)$. The variables $x$ and $y$ denote the left and right leaf, respectively. For each pattern, only the one that minimizes  $S\big( (\TEP,2)_A \big)$  under this pattern is given.}     \label{table:S-TEP2}
	\begin{tabular}{|c|c |c |c |c|c|} 
		\hline
			&	$(A_M,\emptyset,\emptyset)$	&	$(A_M,x,\emptyset)$		&	$(A_M,x,\emptyset)$		&	$(\emptyset,x,y)$		&	$(A_M,x,y)$	 
		\\
		\hline
		$|A| = \ell$	&	$|A_M| = \ell$		&	\thead{$|A_M| + 1 = \ell$ \\$\ell - 1  \le (k-1)k$ \\$\ell-1 = pk+q, 0 \le q < k$}		&	\thead{$|A_M| + 1 = \ell$ \\$\ell - 1  > (k-1)k$}	&	$\ell = 2$	&	\thead{$|A_M| + 2 = \ell$ \\ $\ell \ge 3$}
		\\
		\hline
		Choice of $A_M$		&	arbitrary	&	 \thead{The first $p$ columns \\plus the first $q$ entries \\in the $(p+1)$-th column} 	&	n/a &		irrelevant	&	arbitrary
		\\
		\hline
		$S\big( (\TEP,2)_A \big)=?$	&	$k^\ell$		&	$qk^{p+1} +(k-q)k^p$	&	$\ge k^k$	&	$k^2$	&	$k^2+k + 2 -\ell$
		\\
		\hline
	\end{tabular}
\end{table}

	\begin{table}[ht!]
	\centering
	\caption{The minimizer $A^*$ for $\min_{|A| = \ell} S( (\TEP,2)_A )$.}     \label{table:S-TEP2-minimizer}
	\begin{tabular}{|c|c |c |} 
		\hline
			&	$1\le \ell \le k+1$	&	$k+2 \le \ell \le n_2$	 
		\\
		\hline
		$A^* = ?$		&	$(A_M,x,\emptyset)$		&	$(A_M,x,y)$
		\\
		\hline
		$S\big( (\TEP,2)_{A^*} \big)=?$	&	\thead{$(\ell-1)(k-1)+k$ \\ $=k^2-(k-1)(k+1-\ell)$}		&	\thead{$k^2+k + 2 -\ell$ \\ $=k^2 - (\ell - (k+2))$}
		\\
		\hline
	\end{tabular}
	\end{table}

\begin{theorem}   \label{thm:S-TEP3}
    Let $1 \le \ell \le n_3$ be an integer.
    Let $c \in [0,1)$. Define
    $V_c = k+2$ and 
    $W_c = ck^2$.
    \begin{enumerate}[(1)]
        \item $S(\TEP,3) \le k(k^2 - k + 2) \le k^3$.
	        In particular, 
	            $S(\TEP,3) < k^3$
	        for $k\ge 3$.
	    \item Let $k \ge 2/(1-c)$. If 
                $n_2 + V_c + 3k/2 \le \ell \le n_2 + W_c$,
            then
	            $S((\TEP,3),\ell) \ge (1-c)k^3/8$.
	        In particular,
	            $S(\TEP,3) \ge (1-c)k^3/8$.
    \end{enumerate}
\end{theorem}

\begin{proof}
    (1) To show the upper bound, 
	it suffices to show for every $1 \le \ell \le n_3$, there exists $A \subseteq [n_3]$ of size $|A| = \ell$ such that
	$S((\TEP,3)_A) \le k(k^2 - k + 2)$.
	Let $A= (A_M,A_L,A_R)$. We consider the following cases.
	\begin{enumerate}[(i)]
		\item $2n_2 < \ell \le n_3$. 
			Choose 
				$A= (A_M,L,R)$. 
			Observe that this choice reduces the problem to that height $h=2$ case corresponding to the pattern $(A_M, x, y)$ in Table \ref{table:S-TEP2}. Hence, by Table  \ref{table:S-TEP2},
				$S((\TEP,3)_A) \le k^2 + k - 1$.
		
		\item $n_2+k^2 \le \ell \le 2n_2$. 
		    Equivalently, $\ell = n_2 + k^2$ or $n_2 + k^2 + 1$ or $n_2 + k^2 + 2$. 
		    Choose 
		        $A= (\emptyset,L,A_R)$.
		    Then, 
		        $|A_R| = k^2$ or $k^2 + 1$ or $k^2 + 2$. 
		    By Lemma \ref{lem:A_M-small} and Table \ref{table:S-TEP2-minimizer},
		        \[
		            S((\TEP,3)_A) = S((\TEP,2)_L) \cdot S((\TEP,2)_{A_R})
		            \le k (k+2),
		        \]
		    where the upper bound is achieved at $|A_R| = k^2$. 
		
		\item $n_2 + 2k \le \ell < n_2 + k^2$.	
			Choose
				$A= (\emptyset, L, A_R)$.
			Then, 
			    $2k \le |A_R| < k^2$.
			By Lemma \ref{lem:A_M-small} and Table \ref{table:S-TEP2-minimizer}
			    \[
				    S((\TEP,3)_A) 
				    = S((\TEP,2)_{L}) \cdot S((\TEP,2)_{A_R}) 
				    \le k(k^2 - k + 2),
				\]
			where the upper bound is achieved at $|A_R| = 2k$.
		
		\item $n_2 < \ell \le n_2 + 2k - 1$.
		 	Choose
				$A= (A_M,L,\emptyset)$.
			Observe that this choice reduces the problem to height $h=2$ case corresponding to the pattern $(A_M, x,\emptyset)$ in Table \ref{table:S-TEP2}. 
			Also 
				$1\le |A_M| = \ell - n_2 \le 2k-1$. 
			By Table \ref{table:S-TEP2}, there exists $A_M$, such that 
				$S((\TEP,3)_{A}) \le (k-1)k^2 + k = k(k^2 - k + 1)$,
			where the upper bound is achieved at $|A_M| = 2k-1$.
			
		\item $1 \le \ell \le n_2$. 
			Choose
				$A= (\emptyset,A_L,\emptyset)$. 
			By Lemma \ref{lem:A_M-small} and Theorem \ref{thm:S-TEP2}, 
				$S((\TEP,3)_A) = S((\TEP,2)_{A_L}) \le k^2$.
	\end{enumerate}
	To sum up, in all cases there exists $A$ such that
    	$S((\TEP,3)_A) \le k(k^2 - k + 2)$
	as desired.

\smallskip

(2) Let 
	    $A=(A_M, A_L, A_R) \subseteq [n_3]$
	be such that 
	    $|A_M| + |A_R| + |A_R| = \ell$  
	where
	    $n_2 + V_c + 3k/2 \le \ell \le n_2 + W_c$.
	We show 
	    $S\big( (\TEP,3)_A \big) \ge \Omega(k^3)$.
	\begin{enumerate}[(i)]
		\item $|A_M| \le k$. 
		    Let $|A_M| = \ell'$. Then, 
			    $0 \le \ell' \le k$.
			Let 
				$a = |A_L|$ and 
				$b = |A_R|$.
			Hence, 
			    \[
			        a+b = \ell - |A_M| \ge n_2 + V_c + 3k/2 - \ell' \ge n_2 + V_c + k/2.
			    \]
			Since 
			    $a,b \le n_2$,
			one has 
			    $\min\{a, b\} \ge V_c + k/2 \ge V_c$.
			Hence, by Table \ref{table:S-TEP2-minimizer}, 
			$S((\TEP,2),a) = k^2 + k + 2 - a$ and 
			$S((\TEP,2),b) = k^2 + k + 2 - b$.   
			Since 
			    $a+b = \ell - |A_M| \le n_2 + W_c$,
			by Lemma \ref{lem:A_M-small}, 
				\begin{align*}
				    S((\TEP,3)_A) 
    				&\ge S((\TEP,2)_{A_L}) \cdot S((\TEP,2)_{A_R})  \\
    				&\ge S((\TEP,2), a) \cdot S((\TEP,2), b)  \\
	    			&\ge k \cdot (1-c)k^2 = (1-c)k^3
				\end{align*}
			minimized at $a = n_2, b=W_c$.

		\item $|A_M| \ge k+1$ and $|A_L|, |A_R| \le n_2 - 1$.
			By Lemma \ref{lem:easy-case},
			    $S((\TEP,3)_A) 
				\ge k^{|A_M|} \ge k^{k+1}$.
				
		\item $|A_M| \ge k+1$ and $|A_L| = n_2$.
		
		    For every $i \in [k]$, let 
    		    $r_i$
    		denote the number of entries of $A_M$ in row $i$. 
			\begin{itemize}
				\item $r_i \ge 3$ for some $i$. 
					By Lemma \ref{lem:one-row},
					    $S((\TEP,3)_A) \ge k^3$.
						
				\item $r_i \le 2$ for every $i\in [k]$, and 
				        $|A_M| \ge 3k/2$.
				    
    			    For $t=0,1,2$, let 
    			        $p_t = |\{r_i: r_i=t\}|$.
    			    Then,
    			        \[
    			            p_0 + p_1 + p_2 = k, \quad
    			            p_1 + 2p_2 = |A_M| \ge 3k/2.
    			        \]
    			    Hence,
                            $k + p_2 \ge (p_1 + p_2) + p_2 \ge 3k/2$,
                        this implies 
    			        $p_2 \ge k/2$.
    			    By Lemma \ref{lem:one-row},
    			        \[
    			            S((\TEP,3)_A) \ge \sum_{i=1}^k k^{r_i} \ge p_2 k^2 \ge k^3/2.
    			        \]
			    \item $r_i \le 2$ for every $i\in [k]$, and 
				        $k+1 \le |A_M| < 3k/2$.
                    Hence,
    				    $V_c  \le |A_R| = \ell - n_2 - |A_M| \le W_c$.
    				    
        		    Use the notation $p_0,p_1,p_2$ from above. 
    				Let 
    				    \[
    				        G(A_M) = \{i \in [k]: r_i = 0, 1\}.
    				    \]
    				Then,
    				    $|G(A_M)| = p_0 + p_1 \ge k/4$.
    				Since 
    				    $A_L = L$,
    				we use the notation 
    				    $\langle i \rangle$
    				to denote
    				    $\langle \alpha_L \rangle$
    				for which
    				    $(\TEP,L)(\alpha_L) = i$.
    				Consider the set of pairs
    				    \[
    				        \{(\langle i \rangle, \langle \alpha_R \rangle):
    				        i \in G(A_M), 
    				        (\TEP,2)_{A_R, \alpha_R} \text{ is a non-constant function}\}.
    				    \]
    				Then, we claim that each such different pair defines a distinct subfunction
    				    $(\TEP,3)_{A,\alpha}$
    				for 
    				    $\alpha = (\alpha^*_M, \langle i \rangle, \langle \alpha_R \rangle)$
    				where 
    				    $\alpha^*_M$
    				can be appropriately chosen.
    				Indeed, since 
    				    $i \in G(A_M)$
    				and 
    				    $(\TEP,2)_{A_R, \alpha_R}$
    				is a non-constant function, there exists 
    				    $\beta_R \in [k]^{R - A_R}$
    				such that
    				    $(i,(\TEP,2)_{A_R, \alpha_R}(\beta_R)) \not\in A_M$.
    				Then, it is not hard to see that each such pair would define a distinct subfunction.
    				By Theorem \ref{thm:S-TEP2}, the number of such pairs is
    				    \[
    				        |G(A_M)| \cdot \big( S((\TEP,2)_{A_R}) - k \big)
    				        \ge k/4 \cdot \big( (1-c)k^2  - k\big ) 
    				        \ge (1-c)k^3/8,
    				    \]
    				as long as $k \ge 2/(1-c)$.
			\end{itemize}
	\end{enumerate}
	To sum up, if $k \ge 2/(1-c)$, then in all cases we have shown
    	$S((\TEP,3), \ell) \ge (1-c)k^3/8$
	as desired.
\end{proof}

\begin{theorem}  \label{thm:S-TEP-h}
    For every $h\ge 3$, 
    \[
    S((\TEP,h), \ell) \le 
    \begin{cases}
        k^{h-1}, &\quad 1\le \ell \le \sum_{i=2}^{h-1} n_i, \\
        k^{h-2}(k^2 - k + 2),  &\quad \sum_{i=2}^{h-1} n_i <  \ell < \sum_{i=2}^{h-1} n_i + k^2, \\
        k^{h-2}(k+2), &\quad \sum_{i=2}^{h-1} n_i + k^2 \le \ell \le 2n_{h-1}, \\
        k^2 + k - 1, &\quad 2n_{h-1} < \ell \le n_h.
    \end{cases}
    \]
    In particular, 
        $S(\TEP,h) \le k^h - k^{h-2}(k-2) < k^h$
    when $k \ge 3$.
\end{theorem}   

\begin{proof}
    We use induction on $h$. The base case $h=3$ has been proved in Theorem \ref{thm:S-TEP3}. Assume now $h\ge 4$ and the claim is true for $h-1$. Let 
        $1 \le \ell \le n_h$. 
    Consider $S((\TEP,h),\ell)$. 
    \begin{enumerate}[(i)]
        \item $1\le \ell \le \sum_{i=2}^{h-1} n_i$.
            When 
                $1 \le \ell \le n_{h-1}$,  
            choose 
                $A=(\emptyset, A_L, \emptyset)$.
            Apply the induction hypothesis and Lemma \ref{lem:A_M-small},
                \[
                    S((\TEP,h)_A) = S((\TEP,h-1)_{A_L}) \le  k^{h-1}.
                \]
            When 
                $n_{h-1} < \ell \le \sum_{i=2}^{h-1} n_i$, 
            choose 
                $A=(\emptyset, L, A_R)$.
            Then, 
                $1 \le |A_R| \le \sum_{i=2}^{h-2} n_i$.
            Apply the induction hypothesis and Lemma \ref{lem:A_M-small},
                \[
                    S((\TEP,h)_A) = S((\TEP,h-1)_L) \cdot S((\TEP,h-1)_{A_R}) \le k \cdot k^{h-2} = k^{h-1}.
                \]
        
        \item $\sum_{i=2}^{h-1} n_i <  \ell < \sum_{i=2}^{h-1} n_i + k^2$. 
            Choose 
                $A=(\emptyset, L, A_R)$.
            Then, 
                $\sum_{i=2}^{h-2} n_i < |A_R| < \sum_{i=2}^{h-2} n_i + k^2$.
            Apply the induction hypothesis and Lemma \ref{lem:A_M-small},
                \[
                    S((\TEP,h)_A) = S((\TEP,h-1)_L) \cdot S((\TEP,h-1)_{A_R}) \le k \cdot k^{h-3}(k^2-k+2) = k^{h-2}(k^2-k+2).
                \]
        
        \item $\sum_{i=2}^{h-1} n_i + k^2 \le \ell \le 2n_{h-1}$.
            Choose 
                $A=(\emptyset, L, A_R)$.
            Then, 
                $\sum_{i=2}^{h-2} n_i + k^2 \le |A_R| \le  n_{h-1}$.
            Apply the induction hypothesis and Lemma \ref{lem:A_M-small}, 
                \begin{align*}
                    S((\TEP,h)_A) 
                    &= S((\TEP,h-1)_L) \cdot S((\TEP,h-1)_{A_R}) \\
                    &\le k \cdot \max\{k^{h-3}(k+2), k^2 + k + 1\} = k^{h-2}(k+2), 
                \end{align*}
            where we used $h \ge 4$ in the last step.
        
        \item $2n_{h-1} < \ell \le n_h$. 
            This can be proved in the same way as (i) in the proof of Theorem \ref{thm:S-TEP3}.     \qedhere
    \end{enumerate}
\end{proof}

\end{document}